\documentclass[iop,revtex4]{emulateapj} \newcommand{\ltab}{\LongTables}

\usepackage{hyperref,enumerate}
\input{colordvi}

\hypersetup{
    bookmarks=true,                 
    unicode=false,                  
    pdftoolbar=true,                
    pdfmenubar=true,                
    pdffitwindow=true,              
    pdfstartview={FitH},            
    pdftitle={SEGUE high-res},      
    pdfauthor={vmplacco},           
    pdfsubject={Astronomy},         
    pdfcreator={dvipdf},            
    pdfproducer={dvipdf},           
    pdfkeywords={metal-poor stars}, 
    pdfnewwindow=true,              
    colorlinks=true,                
    linkcolor=red,                  
    citecolor=blue,                 
    filecolor=magenta,              
    urlcolor=cyan,                  
    breaklinks=true,
    linktocpage
}

\newcommand{\eps}[1]{\ensuremath{\log\epsilon\,(\mathrm{#1})}}
\newcommand{\xx}{{\tablenotemark{a}}}
\newcommand{\yy}{{\tablenotemark{b}}}

\newcommand{\ump}{\object{SDSS~J1204$+$1201}}
\newcommand{\emp}{\object{SDSS~J1322$+$0123}}
\newcommand{\caf}{\object{SDSS~J1029$+$1729}}
\newcommand{\abund}[2]{\ensuremath{[\mathrm{#1}/\mathrm{#2}]}}
\newcommand{\cfe}{\abund{C}{Fe}}
\newcommand{\nfe}{\abund{N}{Fe}}
\newcommand{\xfe}[1]{\abund{#1}{Fe}}
\newcommand{\metal}{\abund{Fe}{H}}

\newcommand{\teff}{\ensuremath{T_\mathrm{eff}}}
\newcommand{\logg}{\ensuremath{\log\,g}}
\newcommand{\Msun}{\mathrm{M}_\odot}
\newcommand{\Lsun}{\mathrm{L}_\odot}
\newcommand{\A}[1]{\ensuremath{A(\mathrm{#1})}}
\newcommand{\dex}{\ensuremath{\mathrm{dex}}}
\newcommand{\erg}{\ensuremath{\mathrm{erg}}}
\newcommand{\K}{\ensuremath{\mathrm{K}}}
\newcommand{\Gyr}{\ensuremath{\mathrm{Gyr}}}

\shorttitle{Metal-Poor Stars Observed with the Magellan Telescope. III.}
\shortauthors{Placco et al.}

\begin{document}

\title{Metal-Poor Stars Observed with the Magellan Telescope. III.\footnotemark[1]\\
New Extremely and Ultra Metal-Poor Stars from SDSS/SEGUE and\\
Insights on the Formation of Ultra Metal-Poor Stars}

\footnotetext[1]{Based on observations gathered with the $6.5\,$m Magellan
Telescopes located at Las Campanas Observatory, Chile}

\author{
Vinicius  M.\ Placco\altaffilmark{2},
Anna          Frebel\altaffilmark{3},
Young        Sun Lee\altaffilmark{4},
Heather R.\ Jacobson\altaffilmark{3},\\
Timothy    C.\ Beers\altaffilmark{2},
Jose        M.\ Pena\altaffilmark{3},
Conrad          Chan\altaffilmark{5},
Alexander      Heger\altaffilmark{5,6,7}}

\altaffiltext{2}{Department of Physics and JINA Center for the Evolution of the
                 Elements, University of Notre Dame, Notre Dame, IN 46556, USA}
\altaffiltext{3}{Department of Physics and Kavli Institute for Astrophysics
                 and Space Research, Massachusetts Institute of
                 Technology, Cambridge, MA 02139, USA}
\altaffiltext{4}{Department of Astronomy and Space Science, Chungnam National
                 University, Daejeon 305-764, Republic of Korea}
\altaffiltext{5}{Monash Centre for Astrophysics, School of Physics and Astronomy,
                 19 Rainforest Walk, Monash University, Vic 3800, Australia}
\altaffiltext{6}{Shanghai Jiao-Tong University, CNA, Department of
                 Physics and Astronomy, Shanghai 200240, P.~R.~China}
\altaffiltext{7}{University of Minnesota, School of Physics and Astronomy,
                 Minneapolis, MN 55455, USA}

\addtocounter{footnote}{7}

\begin{abstract}

We report the discovery of one extremely metal-poor (EMP; \metal$<-3$) 
and one ultra metal-poor (UMP; \metal$<-4$) star selected from the SDSS/SEGUE
survey.  These stars were identified as EMP candidates based on their
medium-resolution ($R~\sim2,000$) spectra, and were followed-up with
high-resolution ($R~\sim35,000$) spectroscopy with the Magellan-Clay Telescope.
Their derived chemical abundances exhibit good agreement with those of stars
with similar metallicities.  We also provide new insights on the formation of
the UMP stars, based on comparison with a new set of theoretical models of
supernovae nucleosynthesis.  The models were matched with $20$ UMP stars found
in the literature, together with one of the program stars (\ump), with
$\metal=-4.34$.  From fitting their abundances, we find that the supernovae
progenitors, for stars where carbon and nitrogen are measured, had
masses ranging from $20.5\,\Msun$ to $28\,\Msun$, and explosion energies from
$0.3$ to $0.9\times10^{51}\,\erg$.  These results are highly sensitive to the
carbon and nitrogen abundance determinations, which is one of the main drivers
for future high-resolution follow-up of UMP candidates.  In addition, we are
able to reproduce the different CNO abundance patterns found in UMP stars
with a single progenitor type, by varying its mass and explosion energy.

\end{abstract}


\section{Introduction}
\label{intro}

The most metal-poor stars in the Galactic halo carry important
information about the formation and early evolution of the conditions
in the early universe, as well as in the assembly of the Milky Way.
In particular, ultra metal-poor \citep[UMP;
 \metal\footnote{\abund{A}{B} = $log(N_A/{}N_B)_{\star} -
    \log(N_A/{}N_B) _{\odot}$, where $N$ is the number density of
    atoms of a given element in the star ($\star$) and the Sun
    ($\odot$), respectively.}  $<-4.0$, e.g.,][]{beers2005,frebel2011}
stars are believed to be formed by gas clouds polluted by the chemical
yields of the very first (Population III) stars formed in the universe
\citep{iwamoto2005}.  Even though this scenario for the origin of UMP stars
is qualitatively widely accepted, there are still many open questions,
such as the range of masses and specific characteristics of the
population of first stars, and how many progenitors each UMP star
might have had \citep{tominaga2014}.

There are different scenarios for the progenitor populations that
have been proposed to provide the necessary ingredients to trigger the formation of
UMP stars: (i) pair-instability supernovae from very massive
stars \citep[e.g.,][]{heger2002}, however, the predicted abundance
pattern from such supernova has not been uniquely observed;
(ii) relatively normal massive stars, however, with reduced
mixing relative to their modern counterparts
\citep[e.g.,][]{heger2010}; (iii) Fast-rotating massive stars
\citep{meynet2006,meynet2010} and; (iv) ``Faint'' supernovae
\citep{nomoto2006,nomoto2013,keller2014}.  Analyses of the abundances
of UMP stars, however, are required to ultimately establish the nature
and shape of the initial mass function (IMF) of these Population III
stars.  The goal is to constraint the progenitor properties
including their masses \citep[see discussion in ][]{placco2014}.
Presently, however, there are about $20$ UMP stars with
high-resolution spectra available in the literature, and only five
stars known to be hyper metal-poor \citep[HMP;
  $\metal\leq-5.0$,][]{frebel2015,bonifacio2015,christlieb2002,frebel2005,keller2014}.

Since the completion of the Sloan Extension for Galactic Understanding
and Exploration \citep[SEGUE-1;][]{yanny2009} and SEGUE-2
\citep{rokosi2015}, the numbers of identified Very/Extremely
metal-poor (VMP/EMP, with $\mbox{[Fe/H]}<-2.0$ and $<-3.0$,
respectively) stars have increased by over an order of magnitude, when
compared to previous efforts, such as the HK
\citep{beers1985,beers1992} and Hamburg/ESO \citep{christlieb2008}
surveys.  The SEGUE-1 and SEGUE-2 campaigns, sub-surveys of the Sloan
Digital Sky Survey \citep[SDSS;][]{york2000}, accomplished this with
medium-resolution ($R~\sim2,000$) spectroscopy.  Currently, there are several tens of
thousands of VMP (and on the order of $1000$ EMP) candidates
identified by these surveys, but just a small fraction have been
studied with high-resolution spectroscopy.  Recent efforts to increase
these numbers using SDSS/SEGUE candidates include studies by
\citet{caffau2011a}, \citet{bonifacio2012,bonifacio2015}, \citet{frebel2015}, and
\citet{aoki2013}, and have already resulted in the discovery of
SDSS~J1029$+$1729, with $\metal=-4.99$ \citep{caffau2011b},
SDSS~J1742$+$2531, with $\metal=-4.80$ and SDSS~J1035$+$0641, with
$\metal<-5.07$ \citep{bonifacio2015}, and SDSS~J1313$+$0019, with $\metal=-5.00$
\citep{allende2015,frebel2015}.

In this paper, we present results from a new selection effort for EMP stars
from the data base of medium-resolution SDSS/SEGUE spectra.  We report the
discovery of another UMP star, SDSS~J120441.39$+$120111.5 (hereafter \ump,
$\metal=-4.34\pm 0.05$) and also an EMP star, SDSS~J132250.60$+$012343.0 (hereafter
\emp, $\metal=-3.64\pm 0.05$), selected from medium-resolution SDSS/SEGUE spectra.
We have carried out a detailed chemical-abundance analysis, as well as a
comparison with data from the literature and theoretical models for Population
III stars, in order to gain insights on the progenitor population(s) of UMP
stars.  This paper is outlined as follows: Section~\ref{secobs} describes the
medium-resolution spectroscopy target selection and high-resolution follow-up
observations, followed by the determinations of the stellar parameters and
chemical abundances in Section~\ref{secatm}.  Section~\ref{seccomp} shows a
comparison between (i) the abundances of the program stars and other literature
data, and (ii) the UMP stars in the literature and a new set of theoretical
models for Pop III stars.  Our conclusions are provided in Section~\ref{final}.

\section{Target Selection and Observations}
\label{secobs}

Our targets were selected from the SEGUE-2 database, to avoid possible overlap
with other high-resolution surveys of metal-poor stars based on SDSS/SEGUE
data.  We applied restrictions to the magnitude ($g<16.8$) and metallicity
($\metal<-3.0$).  Estimates of the stellar atmospheric parameters from
medium-resolution SDSS/SEGUE spectra were obtained using the SEGUE Stellar
Parameter Pipeline \citep[SSPP - see][for a detailed
description]{lee2008a,lee2008b,allende2008, smolin2011,lee2011,lee2013}.  These
restrictions cut down the entire SDSS/SEGUE sample to ten potential UMP stars.
Although there are many more possible EMP star candidates selected from the
SDSS/SEGUE-2, we imposed the magnitude constraint in order to obtain candidates
which can be observed with high signal-to-noise, sufficient for detailed
high-resolution spectral analysis, at reasonable exposure times.  After visual
inspection of the ten candidate spectra (and checking their atmospheric
parameters from the SSPP), two objects were followed-up: \ump\ and \emp, which
were the lowest metallicity candidates.  Table~\ref{candlist} lists basic
information on the program stars.  Table~\ref{obstable} lists the SSPP-derived
\teff, \logg, and \metal, used as first-pass estimates when determining the
parameters for the high-resolution analysis, presented in Section~\ref{secatm}.

\begin{figure}[!ht]
\epsscale{1.20}
\plotone{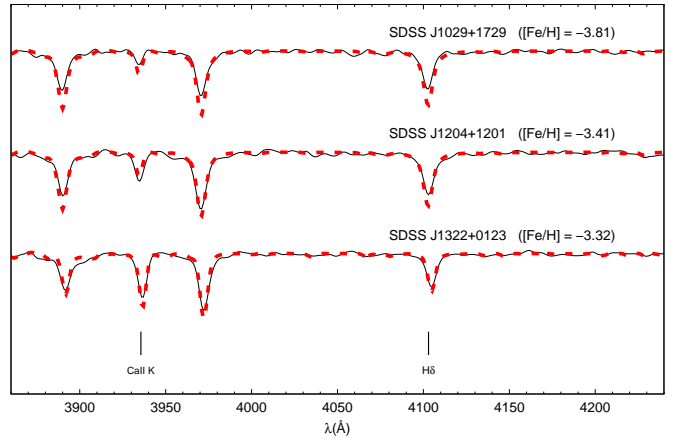}
\caption{Medium-resolution SDSS/SEGUE spectra for the program stars,
  compared with \protect\caf{} (top spectrum), a known UMP star from
  SDSS/SEGUE with $\metal=-4.99$ \citep{caffau2011b}.  The red dashed
  lines show the synthetic spectra from the SSPP.  Also shown are the
  adopted $\metal$ values from the SSPP.}
\label{spec_med}
\end{figure}

Figure~\ref{spec_med} shows the SDSS/SEGUE-2 spectra for the observed stars,
in comparison with the star \caf\ \citep[$\metal=-4.99$ -][]{caffau2011b}, the
first confirmed UMP star from SDSS/SEGUE.  The red dashed lines show the
SSPP spectral templates, used for matching the observed spectra and to
determine the atmospheric parameters.  The adopted SSPP metallicities
are also shown.  It is possible to note a slight mismatch between the
observed and synthetic spectra, in particular for H$\delta$.  This
results in an overestimated temperature, which translates into a
higher $\metal$ value for a given strength of the Ca\,{\sc{ii}}~K
line.  By comparing the SSPP parameters with the high-resolution
determinations (see Section~\ref{secatm} for details), the SSPP
temperatures are overestimated by $\sim400\,\K$, leading to higher
estimated $\metal$.

High-resolution data were obtained during the 2013A semester, using
the Magellan Inamori Kyocera Echelle \citep[MIKE --][]{mike}
spectrograph on the Magellan-Clay Telescope at Las Campanas
Observatory.  The observing setup included a $0\farcs7$ slit with
$2\times2$ on-chip binning, which yielded a resolving power of
$R\sim35,000$ in the blue spectral range and $R\sim28,000$ in the red
spectral range.  The S/N at $5200\,${\AA} is $\sim85$ per pixel (using
integration times of $60$ and $90$ minutes for \emp\ and \ump,
respectively). MIKE spectra have nearly full optical wavelength
coverage over $\sim3500-9000\,${\AA}.  Table~\ref{candlist} lists the
details of the high-resolution observations for the program stars.
The data were reduced using a data reduction pipeline developed for
MIKE spectra, initially described by \citet{kelson2003}\footnote{
  \href{http://code.obs.carnegiescience.edu/python}
       {http://code.obs.carnegiescience.edu/python}}.

\begin{figure}[!ht]
\epsscale{1.20}
\plotone{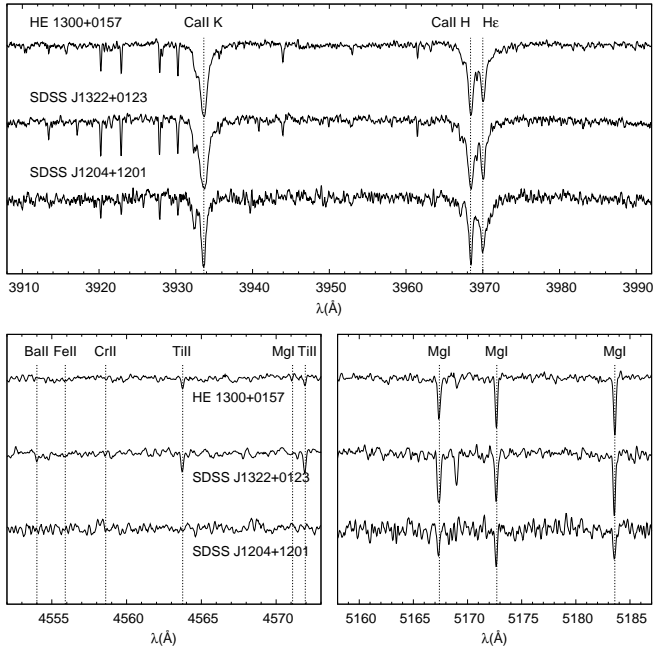}
\caption{Examples of the spectral regions of our program stars around 
  the \ion{Ba}{2} line
  at $4554\,${\AA}, \ion{Mg}{1} triplet, and \ion{Ca}{2} H and K,
  compared with the previously studied star HE~1300$+$0157.}
\label{highres}
\end{figure}

Figure~\ref{highres} shows regions of the MIKE spectra for the program stars in
comparison with the EMP star HE~1300$+$0157, which has similar stellar
parameters as our program stars \citep[$\teff=5450\,\K$, $\logg=3.20$,
$\metal=-3.88$;][]{frebel2007}.  The lower left panel shows the region around
the \ion{Ba}{2} line at $4554\,${\AA}; the lower right panel shows the Mg
triplet around $5170\,${\AA}; and the upper panel shows the \ion{Ca}{2} lines.

\section{Stellar Parameters and Chemical Abundances}
\label{secatm}

\subsection{Techniques}

Atomic absorption lines were measured using the same line list as in
\citep{frebel2014}, based on lines from \citet{aoki2002}, \citet{barklem2005},
and the VALD database \citep{vald}.  Equivalent-width measurements were
obtained by fitting Gaussian profiles to the observed absorption lines.
Table~\ref{eqw} lists the lines used in this work, their measured equivalent
widths, and the derived abundance from each line.

For the abundance analysis, we employ one-dimensional plane-parallel
model atmospheres with no overshooting \citep{castelli2004}, computed
under the assumption of local thermodynamic equilibrium (LTE). The
2011 version of the MOOG synthesis code \citep{sneden1973} was used
for the spectral synthesis. In this version, scattering is treated
with a source function that sums both absorption and scattering
components, rather than treating continuous scattering as true
absorption \citep{sobeck2011}.

Elemental abundance ratios, \xfe{X}, are calculated taking solar
abundances from \citet{asplund2009}.  The average measurements (or
upper limits) for $16$ elements, derived from the MIKE spectra, are
listed in Table~\ref{abund}.  The $\sigma$ values are the standard
error of the mean.  
The abundance uncertainties, as well as the
systematic uncertainties in the abundance estimates due to the
atmospheric parameters, were treated in the same way as described in
\citet{placco2013}.  
For the equivalent-width analysis, any uncertainties calculated to be less 
than the $\sigma$ for \ion{Fe}{1} was replaced by 0.05~dex. For the spectral 
synthesis, a best value for the abundance of a given line is assumed, then 
lower and upper abundance values are set so they enclose the entire spectral
feature. That is taken as the uncertainty. For the systematic uncertainties,
 Table~\ref{sys} shows how changes in each
atmospheric parameter affect the determined abundances.  Also given is
the total uncertainty for each element.

\subsection{Stellar Parameters}

From the high-resolution MIKE spectra, effective temperatures of the
stars were determined by minimizing trends between the abundances derived from
\ion{Fe}{1} lines and their excitation potentials.  The temperatures derived by
this procedure are known to be underestimated when compared with $\teff$ based
on photometry.  As a consequence, such differences also lead to small changes
in surface gravities and chemical abundances.  \citet{frebel2013} provide a
simple linear relation to correct the spectroscopy-derived ``excitation
temperatures'' to photometric-based temperatures. We apply this procedure and
use the corrected $\teff$ to obtain our final stellar parameters.  For warmer
stars on the subgiant branch and near the main-sequence turnoff, these
corrections are fortunately small.  In our cases, the temperature corrections
were $137\,\K$ for \emp, and $188\,\K$ for \ump.  Microturbulent velocities
were determined by minimizing the trend between the abundances of \ion{Fe}{1}
lines and their reduced equivalent widths.

For \emp, the surface gravity was determined from the balance of two
ionization stages for iron lines (\ion{Fe}{1} and \ion{Fe}{2}). We
allowed the difference between the abundances of the \ion{Fe}{1} and
\ion{Fe}{2} lines to be $0.02\,\dex$.  For \ump, since no \ion{Fe}{2}
lines could be measured, the surface gravity was estimated from a
Yale-Yonsei isochrone \citep{demarque2004} with $12\,\Gyr$,
$\metal=-3.5$, and $\xfe{\alpha}=+0.4$.  The final atmospheric
parameters for the program stars are listed in Table~\ref{obstable}.

\subsection{Abundances and Upper Limits}
\label{secab}

\subsubsection{Lithium}

We were able to determine the LTE lithium abundance for \ump.  From the 
doublet at $6707.8\,${\AA}, the spectral synthesis resulted in
$\A{Li}=1.70$\footnote{Here we employ the notation $\A{X}=\eps{X}+12.0$}.  
\citet{lind2009} provide NLTE corrections for Li abundances, based on
evolutionary status and $\A{Li}$. From their Figures 1-3, it is possible to see
that the corrections (for $\teff=6000\,\K$ and $\A{Li}=1.70$) are less than
$0.05\,\dex$, which is within the uncertainty of the observed abundances for
\ump.

\begin{figure}[!ht]
\epsscale{1.20}
\plotone{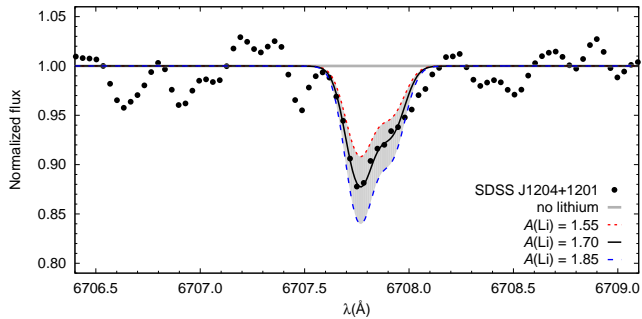}
\caption{Li abundance determination for \protect\ump.  The dots
  represent the observed spectrum, the solid line is the best
  abundance fit, and the dotted and dashed lines indicate the
  abundance uncertainty. The shaded area encompasses a $\pm0.3\,\dex$
  difference in $\A{Li}$.  The light gray line shows the synthesized
  spectrum in the absence of any Li.}
\label{li6707_32}
\end{figure}

Figure~\ref{li6707_32} shows the spectral
synthesis for the Li line in \ump. The dots represent the observed
spectrum, and the solid line is the best abundance fit.  The shaded
area encompasses a $\pm0.3\,\dex$ difference in $\A{Li}$.  The
synthesized spectrum without Li is represented by the light gray line.

\subsubsection{Carbon}

\begin{figure}[!ht]
\epsscale{1.20}
\plotone{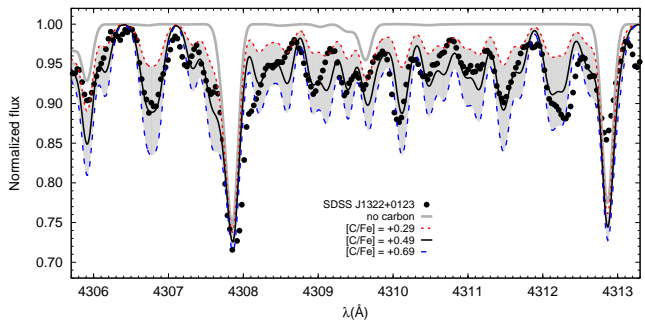}
\caption{Carbon abundance estimate for \protect\emp.  The large black dots
  represent the observed spectrum, the solid line is the best
  abundance fit, and the dotted and dashed lines indicate the
  abundance uncertainty. The shaded area encompasses a $\pm0.2\,\dex$
  difference in \cfe.  The light gray line shows the synthesized
  spectrum in the absence of any carbon.}
\label{ch4313_33}
\end{figure}

It was possible to measure the carbon abundance for \emp\ ($\cfe=+0.49$), using
the CH $G$-band region around $4300\,${\AA}.  For \ump, an upper limit was
determined ($\cfe<+1.45$), using the procedure described in \citet{frebel2006}.
Figure~\ref{ch4313_33} shows the CH $G$-band spectral synthesis for \emp.  The
large black dots represent the observed spectrum, and the solid line
is the best abundance fit.  The dotted and dashed line represent a
$\pm0.2\,\dex$ variation in \cfe, which we conservatively use as the
uncertainty.
We also calculated the carbon abundance corrections for \emp, based on the
procedure described by \citet{placco2014c}.  For $\logg=1.95$, $\metal=-3.64$,
and $\cfe=+0.49$, the correction is $0.01\,\dex$ for the $\xfe{N}=0.0$ case.

\subsubsection{From Na to Ni}

For the elements Na, Mg, Al, Si, Ca, Sc, Ti, Cr, Mn, Co, and Ni,
abundances were determined from equivalent-width analysis only.  For
\emp, no particular discrepancies were found for elements with
abundances determined for more than one line.  The standard errors of
the average abundances are typically smaller than $0.10\,\dex$. In the
case of \ump, only Na and Mg (besides Fe) have more than one measured
line, both with $\sigma=0.05\,\dex$.

\subsubsection{Neutron-capture Elements}

\begin{figure}[!ht]
\epsscale{1.20}
\plotone{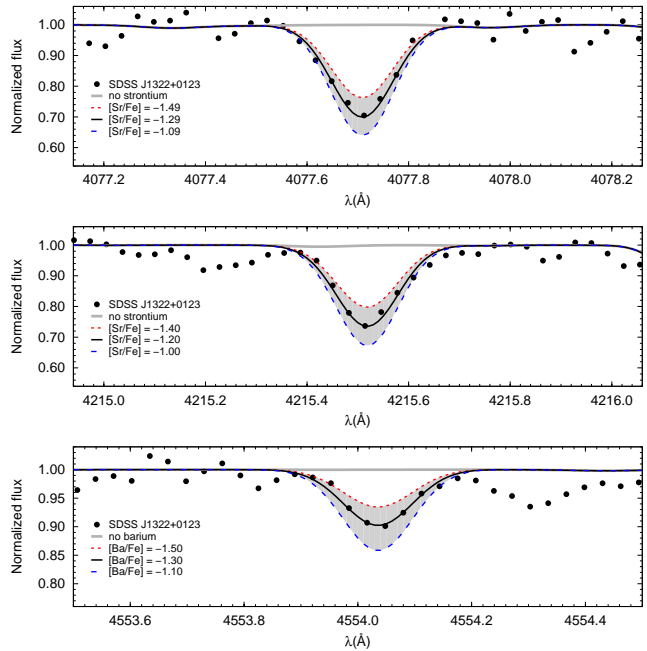}
\caption{Sr and Ba abundance determinations for \protect\emp.  The
  dots represent the observed spectrum, the solid line is the best
  abundance fit, and the dotted and dashed lines indicate the
  abundance uncertainty. The shaded area encompasses a $\pm0.2\,\dex$
  difference in \eps{Sr} and \eps{Ba} abundances.  The light gray line
  shows the synthesized spectrum in the absence of any Sr and Ba,
  respectively.}
\label{basr_syn33}
\end{figure}

For \emp, neutron-capture element abundances were determined from spectral
synthesis.  For \ion{Sr}{2}, the lines used were 4077\,{\AA}
(\abund{Sr}{Fe}=$-1.29$) and $4215\,${\AA} ($\abund{Sr}{Fe}=-1.20$); the
\ion{Ba}{2} abundance was measured from the $4554\,${\AA}
($\abund{Ba}{Fe}=-1.30$) line.  Figure~\ref{basr_syn33} shows the comparison
between the observed and synthetic spectra for these three lines in \emp.
There is good agreement between the abundances of the two Sr lines (values
within $0.1\,\dex$). Only upper limits were determined for \ump\
($\abund{Sr}{Fe}<-0.08$ and $\abund{Ba}{Fe}<+0.62$).

\section{Discussion}
\label{seccomp}

There are many reasons why observations of new stars in the
metallicity range $\metal<-3.5$ are important.  These include a proper
description of the low-metallicity tail of the Galactic Halo MDF, as
well as the nature of the progenitor populations of UMP stars.  Below
we show a comparison between the abundances of the program stars and
other stellar abundances from the literature, as well as an attempt to
describe the main characteristics (e.g., mass and explosion energy) of
UMP star progenitors.

\subsection{Comparison with Literature Abundance Trends}

\subsubsection{Lithium}

\begin{figure}[!ht]
\epsscale{1.20}
\plotone{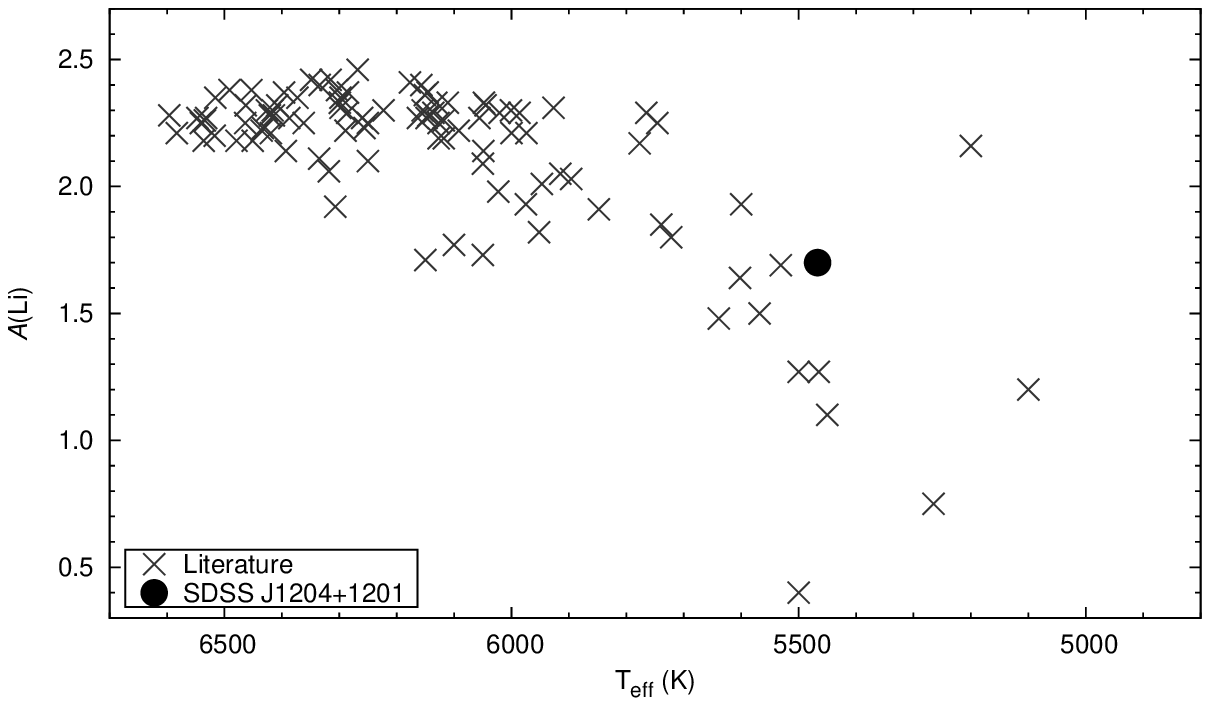}
\caption{$\A{Li}$ vs. $\teff$ for \protect\ump, compared to literature
  data.  References include
  \citet{frebel2007,melendez2010,masseron2012,hansen2014,hansen2015}.}
\label{licomp}
\end{figure}

\ump{} is a subgiant ($\log(L/\Lsun)=1.04$, assuming
$M=0.8\,\Msun$), and has $\A{Li}=1.7$, well below the Li plateau for
metal-poor dwarf stars described by \citet{spite1982}.
Figure~\ref{licomp} shows the behavior of Li abundances, as a function
of temperature, for a sample of metal-poor stars from the
literature. Individual references are listed in the figure caption.
The Li $\A{Li}=1.7$ abundance for \ump{} is consistent with data from
the literature at the same temperature and metallicity, and suggests
that Li is already being depleting on the lower giant branch, due,
at least in part, to internal processing.

\subsubsection{Other Elements}

\begin{figure}[!ht]
\epsscale{1.20}
\plotone{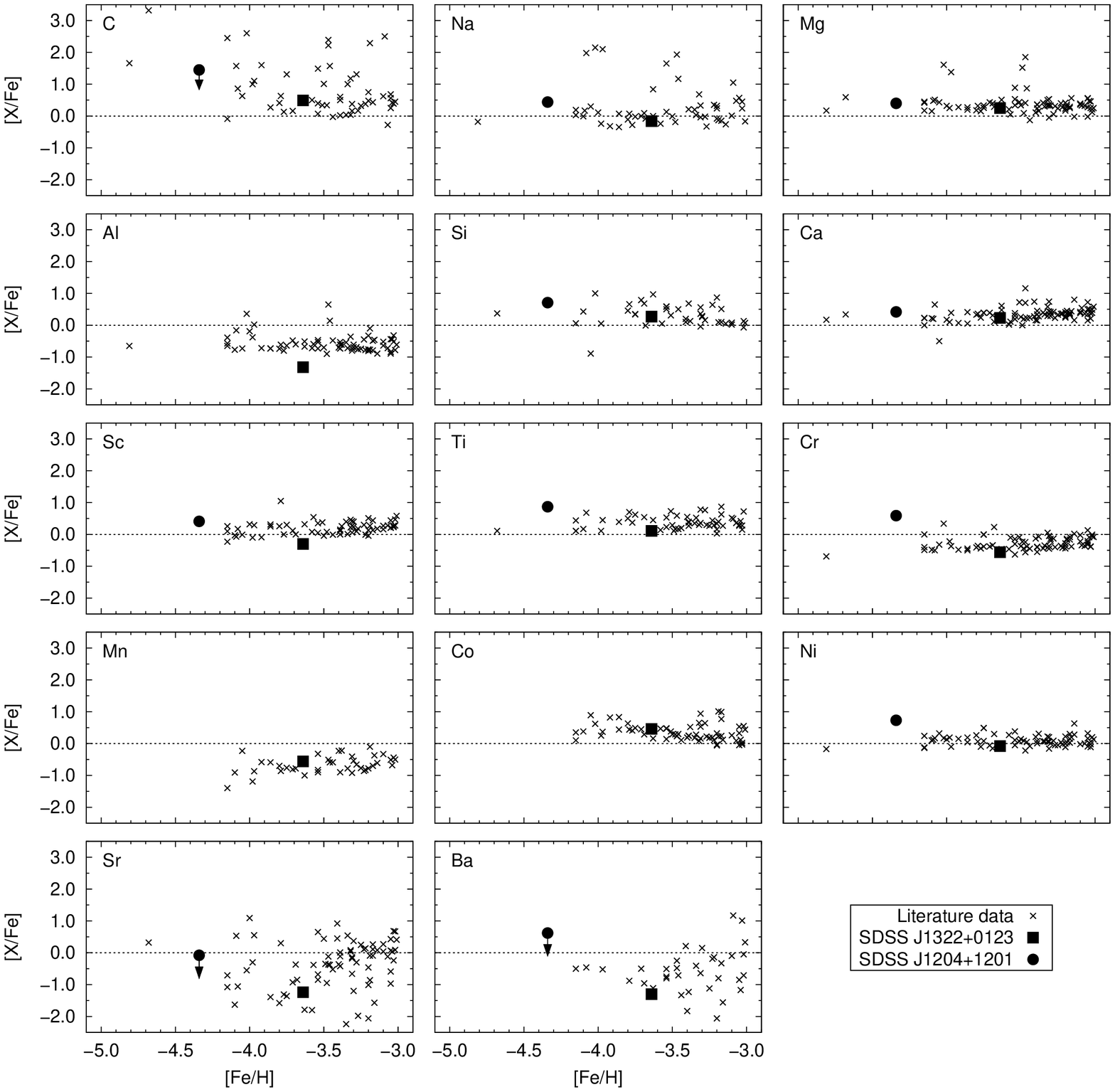}
\caption{$\xfe{X}$ vs. $\metal$ for carbon, $\alpha$-, iron-peak, and
  neutron-capture elements, for the program stars and stars with
  $\metal<-3.0$ from \citet{yong2013} and \citet{hansen2014}.}
\label{abfull}
\end{figure}

We also compared the abundances of our sample stars with the EMP star
samples from \citet{yong2013} (giants only) and
\citet{hansen2014}.  Results are shown in Figure~\ref{abfull}.  The
carbon abundances from the literature were also corrected using the
procedure described in \citet{placco2014c}.  No significant differences
are found for the light elements, except for Al, which falls below the
trend for the literature data.  The upper limits of C, Sr, and Ba for
\ump\ are also consistent with typical values for halo stars in this
metallicity range.

{\emp} is moderately enhanced in carbon ($\cfe=+0.49$).  Given its high
gravity, the carbon correction is only $0.01\,\dex$, leading to
$\cfe_c=+0.50$.  Even though in the $\metal<-3.5$ metallicity range,
$70\,\%$ of the stars exhibit $\cfe>+0.50$ \citep{placco2014c}, the
carbon abundance of \emp{} appears to be within expectations.  For
\ump, the $\cfe<+1.45$ is also within expectations, given that
$80\,\%$ of the stars with $\metal<-4.3$ show $\cfe>+1.0$.  However, 
a higher S/N spectrum is needed for a proper carbon abundance determination 
for \ump. The abundances of Ti, Cr, and Ni for \ump{} appear to be higher
than the trends presented by the literature data. We caution the reader,
however, that the abundances for these species were derived by equivalent-width
analysis of only one spectral feature. Further measurements are needed to
properly address this issue.

In the $\metal<-3.0$ regime, the neutron-capture elements Sr and Ba
are thought to be formed by the $r$-process, rather than the
$s$-process, which dominates at $\metal>-2.5$ \citep{placco2013}.
The low values determined for \emp{} are consistent
with the literature data.  The $\abund{Sr}{Ba}=+0.06$ value is also
within the range for stars with $\metal<-3.5$ \citep{aoki2013}.

\subsection{Model Predictions for UMP Progenitors}
\label{fitsec}

In this section we attempt to assess the properties (mass distribution and
explosion energies) of the progenitor population of UMP stars.  For this
exercise, we used \ump{} and $19$ UMP stars with parameters and abundances
determined from high-resolution spectroscopy, gathered by \citet{placco2014c}.
Individual references and parameters are listed in Table~\ref{umptab}.  The
abundance data were compared to the theoretical model predictions for
non-rotating single massive Population III stars, in the range
$10\,\Msun-100\,\Msun$ of \citet{heger2010}.  The model database comprises
$120$ initial masses, and explores a range of explosion energies from
$0.3\times10^{51}\,\erg$ to $10\times10^{51}\,\erg$ kinetic energy of the
ejecta.  The models further vary the amount of mixing in the supernova ejecta
due to Rayleigh Taylor instabilities \citep[e.g.,][]{joggerst2009}.  Here we
also use their $\chi^2$ matching algorithm \citep{heger2010}.  An online tool
including the model database can be found at
\texttt{starfit}\footnote{\href{http://starfit.org}{http://starfit.org}}.  The
same fitting procedure was also used in \citet{keller2014}, \citet{bessell2015},
and \citet{frebel2015}.  In this work we adopt a similar procedure as the one
described in \citet{heger2010}, using all available element measurements and
upper limits up to atomic number $Z=30$.  For Cr and Sc, the authors assumed
that there are additional production sites not included in the model data, e.g.,
contributions from neutron star winds, and hence the model yield was taken as
lower limit.

\subsubsection{Best Model Fits}

Chemical abundances (and/or upper limits) for the UMP sample are
available for the following species: C, N, O, Na, Mg, Al, Si, Ca, Sc,
Ti, V, Cr, Mn, Fe, Co, Ni, and Zn. For the solar abundances, the
\texttt{starfit} code uses the photospheric values from
\citet{asplund2009}.  

\begin{figure*}[!ht]
\epsscale{0.80}
\plotone{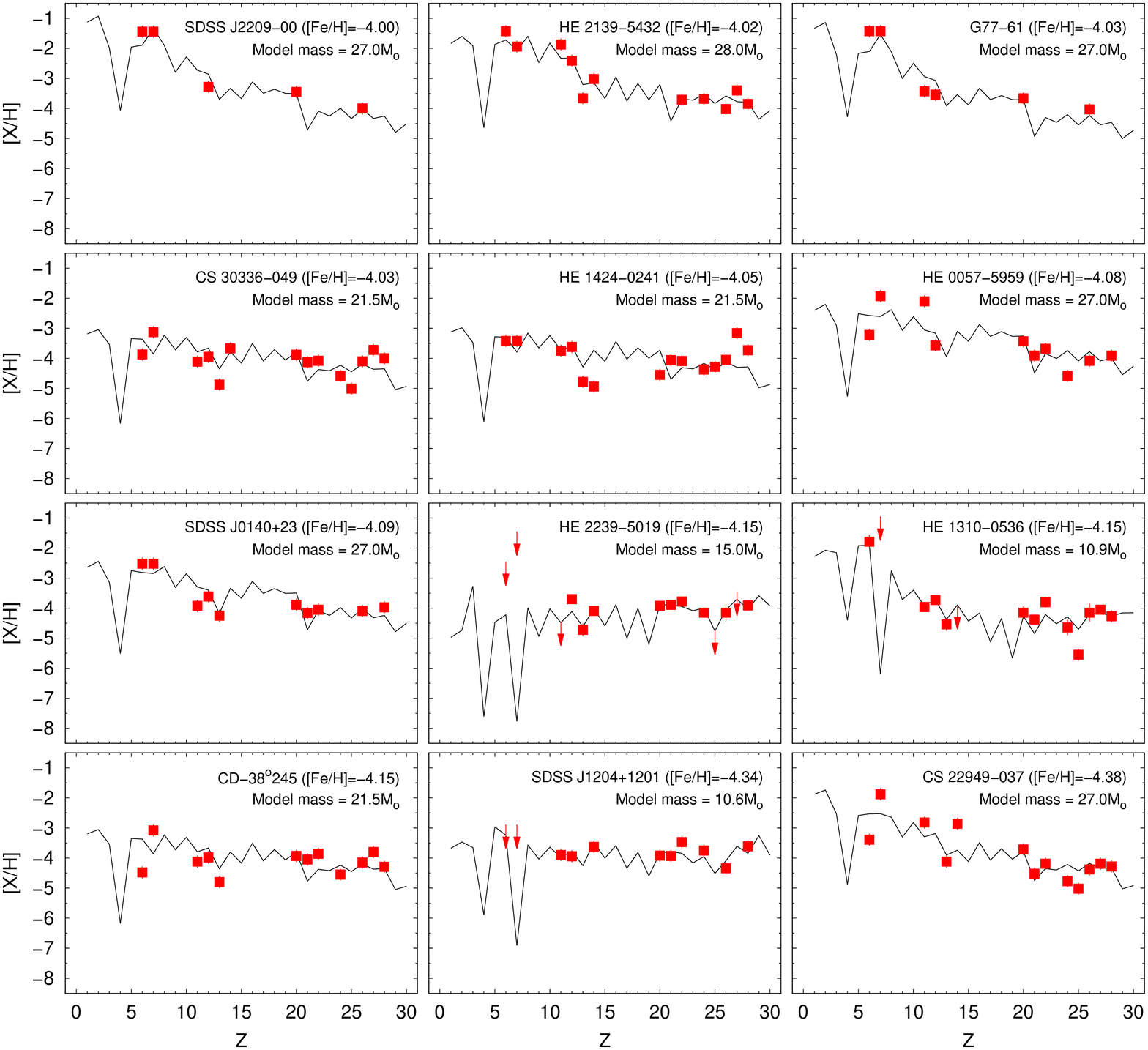}
\caption{Abundance ratios $\abund{X}{H}$ as a function of charge
  number $Z$ for the first twelve UMP stars.  \textsl{Red filled
    squares} are abundances taken from the literature (see text for
  comments on carbon and nitrogen).  The \textsl{solid line} is the
  best fit for each star, with the model mass shown in the upper
  right.  \textsl{Arrows} represent upper limits.}
\label{starfit1}
\end{figure*}

\begin{figure*}[!ht]
\epsscale{0.80}
\plotone{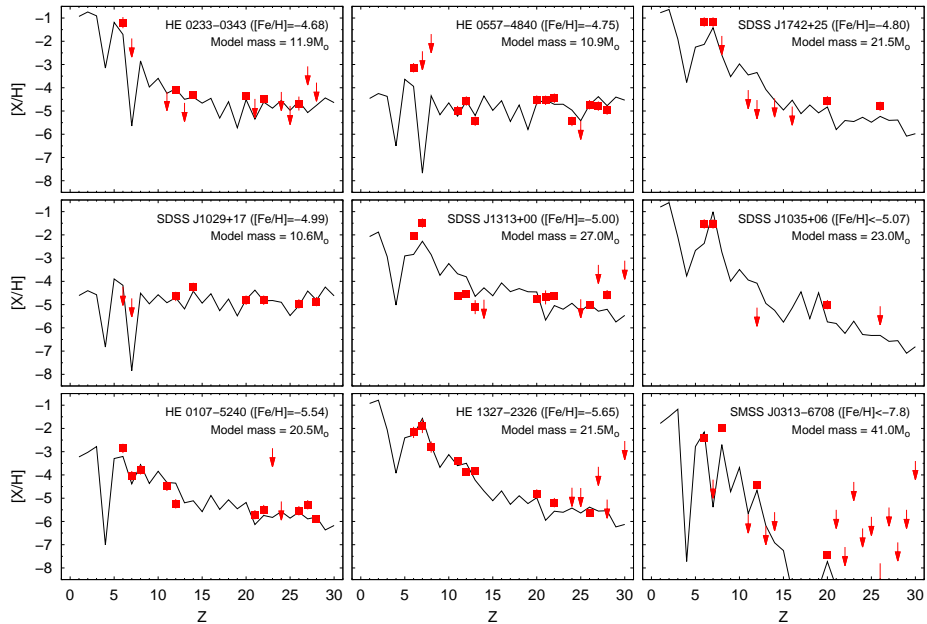}
\caption{Abundance ratios $\abund{X}{H}$ as a function of charge
  number $Z$ for the second nine UMP stars.  \textsl{Red filled
    squares} are abundances taken from the literature (see text for
  comments on carbon and nitrogen).  The \textsl{solid line} is the
  best fit for each star, with the model mass shown in the upper
  right.  \textsl{Arrows} represent upper limits.}
\label{starfit2}
\end{figure*}

Figures~\ref{starfit1} and \ref{starfit2} show
the best model fit for the $21$ UMP stars in Table~\ref{umptab}.  The red
filled squares represent all the abundances gathered from the
individual studies (including upper limits as arrows). The solid black
line is the best model fit to the data. Among the $21$ stars, six do not
have nitrogen abundance measurements or upper limits.  For those, we
estimated the nitrogen abundance using $\abund{C}N=0.0$ \citep[see
  discussion below and top panel of Figure~11 in][]{placco2014c}. For
the carbon abundances, seven stars had their abundances corrected for
their evolutionary status using the procedures described in
\citet{placco2014c}. A summary of the abundances and corrections
applied to the data is shown in Table~\ref{umptab}.

From the figures, one can see no relation between the progenitor mass and the
iron abundance of the UMP stars.  This may be expected, since the
$\metal<-4.0$ range is believed to probe second-generation stars, for which the
local enrichment from the Pop III star explosion is expected to dominate over
global chemical evolution effects.  In addition, it is possible to see that the
models always predict a very low nitrogen abundance ($\abund{N}{H}<-6.0$) where
only upper limits on nitrogen are determined.  This has a direct impact on the
best model chosen by the \texttt{starfit} code, and introduces an artifact on
the progenitor mass distribution, as explained below.
Also, for stars where \abund{N}{H} is higher than \abund{C}{H} by more than
$1.0\,\dex$ (e.g. HE~0057$-$5959, CD$-$38$^o$245, and CS~22949$-$037), the model
is not capable of reproducing the CN pattern, and the overall fit yields a
higher residual value.

\begin{figure}[!ht]
\centering
\epsscale{1.10}
\plottwo{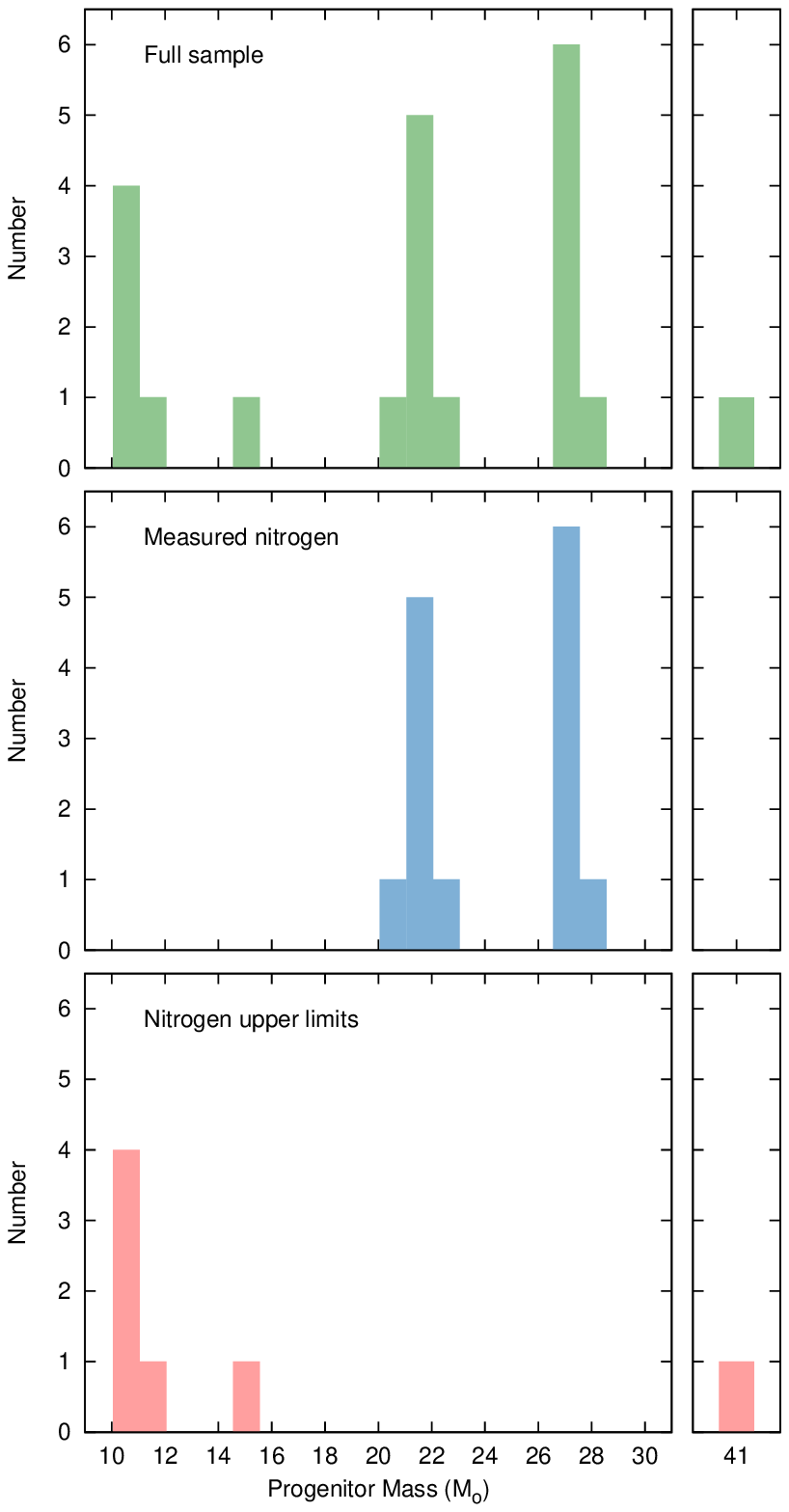}{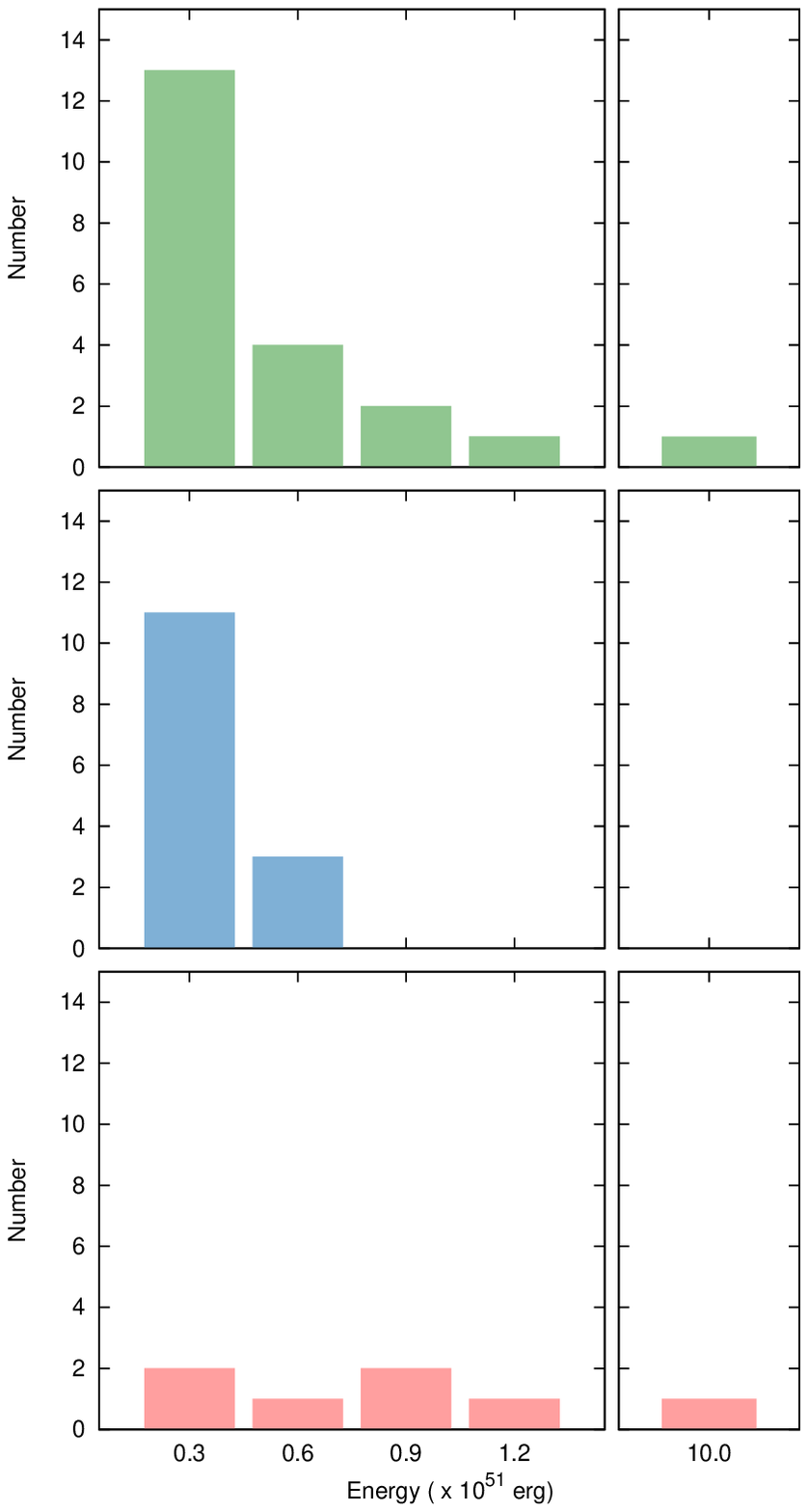}
\caption{Progenitor mass (\textsl{left panels}) and explosion energy
  (\textsl{right panels}) distributions for the $18$ UMP stars.
  \textsl{Upper panels}: full sample.  \textsl{Middle panels}: stars
  with measured nitrogen abundances.  \textsl{Bottom panels}: stars
  with nitrogen upper limits.}
\label{histmass}
\end{figure}

Figure~\ref{histmass} shows the progenitor mass (left panels) and
explosion energy (right panels) distributions for the best model fits.
The upper panels show the full sample ($21$ stars), the middle panels
show stars with measured nitrogen abundances ($14$ stars -- including
the estimates from Table~\ref{umptab}), and the bottom panels show
stars with nitrogen upper limits (seven stars).
For the progenitor mass, a clear separation in the distributions is
seen in the middle and bottom panels.  Progenitor masses for stars with
upper limits on nitrogen concentrate in the $10.6-15.0\,\Msun$
range.  In contrast, progenitor masses for stars with measured nitrogen
abundance show two preferred ranges: $20.5-23.0\,\Msun$ and
$27.0-28.0\,\Msun$.
Concerning the explosion energy, more than $75\,\%$ of the stars with
measured nitrogen show $0.3\times 10^{51}\,\erg$, regardless of the
progenitor mass ranges listed above.  This can either be an indication
of the nature of these progenitors, or an numerical artifact, since
this is the lowest available energy within the model grid.  In one
particular case (HE~2239$-$5019), the explosion energy is the maximum
value allowed by the models ($10\times10^{51}\,\erg$).  This spurious
result could be explained by the lack of carbon and nitrogen abundance
measurements (see explanation below in Sections~\ref{robust} and
\ref{error}.), or by the behavior of the abundance pattern for the
light elements.

For the stars with measured nitrogen, interesting differences in the
progenitor population arise for the lowest metallicity stars in the
sample; one example is the difference in the CNO pattern of
HE~0107$-$5240 and HE~1327$-$2326.  Even though these two stars have
$\metal$ values within $0.2\,\dex$ of one another, and estimated
progenitor masses within $5\,\%$, HE~1327$-$2326 has a $\eps{N}$ two
orders of magnitude higher than HE~0107$-$5240.  In the context of
this work, this difference could be explained by changes in mass and
explosion energy ($0.6\times10^{51}\,\erg$ for HE~0107$-$5240 and
$0.3\times10^{51}\,\erg$ for HE~1327$-$2326), and does not require
additional models to describe the progenitor population.  Another less
extreme comparison is for CS~30336$-$049 and HE~1424$-$0241: their
metallicity, carbon, and nitrogen abundances are within $0.02\,\dex$,
$0.16\,\dex$, and $0.29\,\dex$, respectively, and both share the same
best model fit ($21.5\,\Msun$ and $0.3\times10^{51}\,\erg$).  For
SDSS~J1742$+$25 ($\metal<-5.07$), even though the progenitor mass and
explosion energy ($23.0\,\Msun$ and $0.6\times10^{51}\,\erg$) are
similar to other stars in the same $\metal$ range, the small number of
determined abundances (C and Ca) clearly affects the fitting procedure
(see discussion below).  Additional abundance measurements for these
stars, as well as the discovery of more stars in this \metal{} range
are needed for further investigation.  For the most iron-poor star
known, SMSS~J0313$-$6708, even though our results are consistent with
the models described in \citep{bessell2015}, the lack of nitrogen
abundance measurement is possibly affecting the progenitor mass
determination.

\subsubsection{Robustness of Best Model Fits}
\label{robust}

The \texttt{starfit} procedure gives the ten best model fits for a
given set of input abundances, ranked by their $\chi^2$ value.  To test
the robustness of the best solution, we analyzed the $\chi^2$
variation between the ten best models for each star, and how they
affect the progenitor mass.  For instance, a flat $\chi^2$ distribution
with a wide range of progenitor masses is an indication that the
solutions are not very robust.  Ideally, the $\chi^2$ value should
rapidly increase between the first and second-best solutions.

\begin{figure}[!ht]
\epsscale{1.20}
\plotone{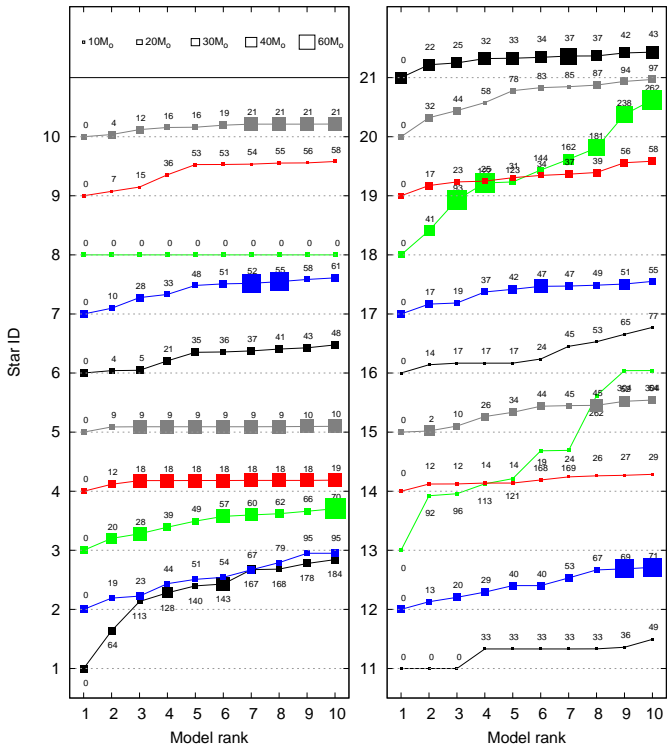}
\caption{Change (in $\chi^2\,\%$) on the residual values as a function
  of the model rank for the sample stars.  The $y$-axis shows the Star
  ID from Table~\ref{umptab}.  The point size is proportional to the
  progenitor mass for each model.}
\label{cases4}
\end{figure}

Figure~\ref{cases4} shows the evolution of the $\chi^2$ values as a
function of the model rank.  Each line represents a star on
Table~\ref{umptab}, labeled by its ``Star ID''.  The numbers above each
point are the variations (in $\%$) of the $\chi^2$ value for a given
model, compared to the $\chi^2$ value of the best model fit.  The point
size is proportional to the progenitor mass, with approximate values
labeled on the top left.
Inspection of Figure~\ref{cases4} reveals that, for $2/3$ of
the sample stars, the variation in $\chi^2$ between the first and
second-best models is above $10\,\%$.  In addition, for six out of
eight stars where the $\chi^2$ variation is below $10\,\%$, the
progenitor masses only changes in two cases. For these stars -- $\#6$
(HE~0057$-$5959) and $\#10$ (CD$-$38~245) -- the variation is between
$21.5\,\Msun$ and $27.0\,\Msun$, which are precisely the two preferred
mass values according to Figure~\ref{histmass} (see discussion above).
For stars $\#4$ (CS~30336$-$049) and $\#5$ (HE~1424$-$0241), there 
is a large change in mass between model-rank 2 and 3, with a small 
variation in $\chi^2$. Even though both stars have 14 determined abundances,
additional work must be performed to evaluate whether the measurements or the
models could be changed to generate a better agreement.
For stars $\#15$ (SDSS~J1742$+$25) and $\#18$ (SDSS~J1035$+$06), the
low number of determined abundances compromises the robustness of the
best model fits.

\subsubsection{Uncertainty/Robustness in the Progenitor Mass}
\label{error}

Chemical abundance measurements carry uncertainties related to the
choice of model atmosphere, continuum placement when measuring
absorption line strengths, and atomic data line lists.  In addition,
non-LTE and 3D effects also have an impact on the determinations.  All
of these changes in the abundances can generate changes in the
progenitor-star properties.  Keeping the observed uncertainty in mind,
we are using C and N as representative examples, as the fitting
results appear to most strongly depend on them.  As shown above, the
existence of a nitrogen measurement is quite important for the model-fit
procedure.

The stellar evolution of primordial stars, which  are the basis of this study,
is very peculiar. These primorial stars have no initial CNO, so in order to
burn hydrogen -- for which they need the CNO cycle -- they need to
produce a trace of CNO material themselves. Typical mass fractions are
$10^{-7}$ or less at the end of core hydrogen burning and in the H-burning shell.
Because of that, any significant amounts of nitrogen found in the UMP stars (and not
made by them \textsl{in situ}) must come from another primordial production
mechanism in the Pop III stars, during or after the onset of central helium
burning, and by mixing of the He burning products with the H envelope.  Without
rotation or other mixing processes, in the models of \citet{heger2010} this
only occurred consistently for initial masses of $45\,\Msun$ and above (their
Figure~11), whereas most of the lower-mass models did not have large nitrogen
yields.  Hence observational constraints from nitrogen measurements or upper
limits can significantly constrain the mass range of models that provide a good
fit.

\begin{figure}[!ht]
\epsscale{1.15}
\plotone{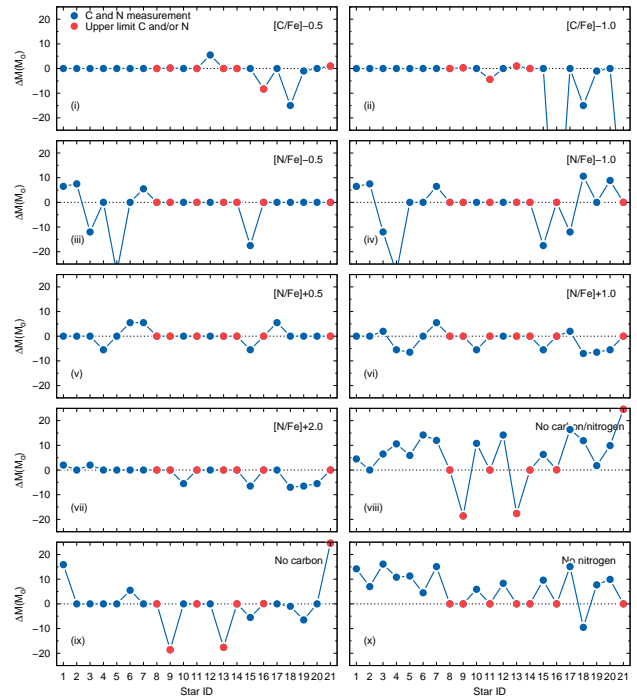}
\caption{Variations (in $\Msun$) in the progenitor mass (compared to
  that of the best-fit model) for the sample stars, for the ten cases
  described above.  \textsl{Blue filled dots} are stars with available
  carbon and nitrogen abundance measurements, and \textsl{red filled
    dots} are stars with upper limits for carbon and/or nitrogen.}
\label{cases2}
\end{figure}

In order to evaluate how changes in the nitrogen (and/or carbon) abundance
reflect on the progenitor mass, we ran the \texttt{starfit} procedure for the
$21$ UMP stars for ten different scenarios, listed in Table~\ref{casestab}.
Then, we evaluated how the progenitor mass of the best model changes.
Figure~\ref{cases2} shows how the progenitor mass changes for each star, for
the ten cases listed in Table~\ref{casestab}, compared with the result of the
best-fit model presented in Figures~\ref{starfit1} and \ref{starfit2}, and
Table~\ref{umptab}.  Blue filled dots are stars with available carbon and
nitrogen measurements, and red filled dots represent stars with available upper
limits for carbon and/or nitrogen.

When changing the carbon abundances -- cases (i) and (ii) -- the stars
affected with available carbon and/or nitrogen measurements are $\#12$ (CS~22949$-$037), 
and $\#18$ (SDSS~J1035$+$0641). Star $\#12$ has its
progenitor mass changed from $27.0\,\Msun$ to $21.5\,\Msun$ (within the mass
range shown in the midle panels of Figure~\ref{histmass}), and star $\#18$
from $23.0\,\Msun$ to $38.0\,\Msun$. The latter case has only three determined
abundances, which makes the model-fitting more susceptible to changes.
For nitrogen, large changes in progenitor mass are seen (for stars
$\#1$, $\#2$, $\#3$, $\#4$, $\#5$, $\#7$, and $\#15$) when reducing the \nfe{}
abundances -- cases (iii) and (iv).  For positive changes in nitrogen -- cases
(v), (vi), and (vii) -- there are smaller progenitor masses variations for a
handful of stars.  These variations, however, are always between
$20\,\Msun-22\,\Msun$ and $27\,\Msun-29\,\Msun$, agreeing with the two
distributions shown in the middle panel of Figure~\ref{histmass}.

When removing the carbon abundances from the fit -- case (ix) -- the
only considerable change (apart from the stars with upper limits) is
for star $\#1$ (SDSS~J2209$-$0028).  This can be explained by the low
number of abundances -- five -- used for the fit.  The most noticeable
changes happen for cases (viii) and (x), when the nitrogen abundance
is removed.  Then, every progenitor mass is changed by at least
$5\,\Msun$.

With these tests, we show that the models are sufficiently robust to
not be affected by typical uncertainties in \cfe{} and \nfe{}
($<0.3\,\dex$).  Moreover, regardless of the number of abundances
available for the fit, we stress the strong effect the presence (or
absence) of nitrogen abundances has on the final result.  Below we
compare our results with a similar study, in which abundance patterns
of stars with $\metal<-3.5$ are compared with a set of SN models with
fixed masses.

\subsubsection{Comparison with \citet{tominaga2014} Results}

\citet{tominaga2014} performed abundance profiling for $48$ stars with
$\metal<-3.5$ from the literature, to gain insight into the
properties of the Pop III progenitors.  The authors use Pop III SN
models with $25\,\Msun$ and $40\,\Msun$.  With the fixed mass,
free parameters are explosion energy, remnant mass, and ejected
mass.  They also include models with $25\,\Msun$ and enhanced mixing
due to rapid rotation.

There are ten stars in common between \citet{tominaga2014} and this
work. For all these stars, the explosion energy of their $25\,\Msun$
SN progenitor is higher (by at least an order of magnitude) then the
values found by this work.  Regardless, it is interesting to see that
the progenitor mass range found by the \texttt{starfit}
($20.5\,\Msun-28.0\,\Msun$ - when using both C and N for the matching)
is consistent with a $25\,\Msun$ progenitor.  For the two most
iron-poor stars analyzed by \citet{tominaga2014} -- HE~0107$-$2326 and
HE~1327$-$2326 -- the authors find similar explosion energies and
remnant masses for both stars, regardless of the different nitrogen
abundances.  In their Figure~7, however, the models cannot
properly reproduce the nitrogen abundance for HE~1327$-$2326.

\citet{ishigaki2014} also performs abundance profiling for stars with
$\metal< -4.5$.  Their best-fit models for HE~0107$-$2326,
HE~1327$-$2326, and HE~0557$-$4840 (high carbon abundances) have
$25\,\Msun$ and explosion energy of $1.0\times10^{51}\,\erg$.  Similar
to what was found by \citet{tominaga2014}, the high N abundance for
HE~1327$-$2326 is also not reproduced, with the models consistently
underproducing nitrogen when compared to the observed values, by at
least $2\,\dex$.  For SDSS~J1029$+$1729, the authors find a suitable
model with $40\,\Msun$ and explosion energy of
$30.0\times10^{51}\,\erg$.  

An alternative formation scenario for this star, which includes dust-induced
cooling, is given by \citet{schneider2012}, where the authors find good
agreement between observations and the yields of core-collapse SNe with metal-free
progenitors of $20\,\Msun$ and $35\,\Msun$.
In the case of
SMSS~J0313$-$6708, models with $25\,\Msun$ and $40\,\Msun$ well-reproduce the
abundances of carbon, magnesium, and calcium, as well as the upper limit for
nitrogen.  In contrast, the model presented in this work for SMSS~J0313$-$6708
also has a good overall abundance fit, even though is has a higher mass
($60\,\Msun$) and lower explosion energy -- $1.0\times10^{51}\,\erg$ as opposed
to $10.0\times10^{51}\,\erg$.

\section{Conclusions}
\label{final}

In this work we presented the high-resolution abundance analysis of
two stars selected from the SDSS/SEGUE survey, with $\metal=-3.64$ and
$\metal=-4.34$.  A detailed chemical abundance analysis reveals that
these stars show the expected behavior of stars in the same
metallicity range.  The addition of these two stars to the
$\metal<-3.5$ range corroborates with the hypothesis presented by
\citet{yong2013b} that the Galactic Halo MDF smoothly decreases down
to $\metal=-4.1$, instead of the sharp cutoff at $\metal=-3.6$
suggested by \citet{schorck2009} and \citet{li2010}.

We also provide new insights on the progenitor population of UMP
stars, by comparing their abundance patterns to a set of theoretical
SN models.  We find that all UMP stars with carbon and nitrogen
abundances available have progenitors with masses in either
$20.5\,\Msun-23.0\,\Msun$ or $27.0\,\Msun-28.0\,\Msun$ ranges, with
explosion energies between $0.3$ and $0.9\times10^{51}\,\erg$.  We
stress that, even though there could be additional suitable candidates
for the UMP progenitors, such as the fast rotating massive stars from
\citet{meynet2006} and the faint supernovae from \citet{tominaga2014},
the models presented in this work are capable of describing the
differences in the abundance patterns by adjusting the progenitor mass
and explosion energy.

Comparison with similar studies from the literature shows that, even
though higher progenitor masses ($M=40\,\Msun$) also show good
agreement for the most extreme cases, such as SMSS~J0313$-$6708,
models have their best fits for $25\,\Msun$, which is consistent with
the mass range found in this work.  Besides having additional targets
at $\metal<-4.0$ for comparison, it is important to have accurate
nitrogen and light element abundances for all UMP stars in the
literature.  There are still a number of similar candidates in need of
high-resolution follow-up, and we expect to conduct similar studies in
the near future.

\acknowledgments

V.M.P.\ and T.C.B.\ acknowledge partial support for this work from
grants PHY 08-22648; Physics Frontier Center/Joint Institute or
Nuclear Astrophysics (JINA), and PHY 14-30152; Physics Frontier
Center/JINA Center for the Evolution of the Elements (JINA-CEE),
awarded by the US National Science Foundation.  
A.F.\ is supported by NSF CAREER grant AST-1255160.  
Y.S.L. acknowledges partial support by the 2014 research fund of 
Chungnam National University and support from the National Research 
Foundation of Korea to the Center for Galaxy Evolution Research.
A.H.\ was supported by a Future Fellowship of the Australian 
Research Council (FT120100363).

\clearpage

\bibliographystyle{apj}

\begin{thebibliography}{}
\expandafter\ifx\csname natexlab\endcsname\relax\def\natexlab#1{#1}\fi

\bibitem[{{Allen} {et~al.}(2012){Allen}, {Ryan}, {Rossi}, {Beers}, \&
  {Tsangarides}}]{allen2012}
{Allen}, D.~M., {Ryan}, S.~G., {Rossi}, S., {Beers}, T.~C., \& {Tsangarides},
  S.~A. 2012, \aap, 548, A34

\bibitem[{{Allende Prieto} {et~al.}(2008){Allende Prieto}, {Sivarani}, {Beers},
  {Lee}, {Koesterke}, {Shetrone}, {Sneden}, {Lambert}, {Wilhelm}, {Rockosi},
  {Lai}, {Yanny}, {Ivans}, {Johnson}, {Aoki}, {Bailer-Jones}, \& {Re
  Fiorentin}}]{allende2008}
{Allende Prieto}, C., {Sivarani}, T., {Beers}, T.~C., {et~al.} 2008, \aj, 136,
  2070

\bibitem[Allende Prieto et al.(2015)]{allende2015} Allende Prieto, 
C., Fernandez-Alvar, E., Aguado, D.~S., et al.\ 2015, arXiv:1505.05555 

\bibitem[{{Aoki} {et~al.}(2002){Aoki}, {Ryan}, {Norris}, {Beers}, {Ando}, \&
  {Tsangarides}}]{aoki2002}
{Aoki}, W., {Ryan}, S.~G., {Norris}, J.~E., {et~al.} 2002, \apj, 580, 1149

\bibitem[{{Aoki} {et~al.}(2013){Aoki}, {Beers}, {Lee}, {Honda}, {Ito},
  {Takada-Hidai}, {Frebel}, {Suda}, {Fujimoto}, {Carollo}, \&
  {Sivarani}}]{aoki2013}
{Aoki}, W., {Beers}, T.~C., {Lee}, Y.~S., {et~al.} 2013, \aj, 145, 13

\bibitem[{{Asplund} {et~al.}(2009){Asplund}, {Grevesse}, {Sauval}, \&
  {Scott}}]{asplund2009}
{Asplund}, M., {Grevesse}, N., {Sauval}, A.~J., \& {Scott}, P. 2009, \araa, 47,
  481

\bibitem[{{Barklem} {et~al.}(2005){Barklem}, {Christlieb}, {Beers}, {Hill},
  {Bessell}, {Holmberg}, {Marsteller}, {Rossi}, {Zickgraf}, \&
  {Reimers}}]{barklem2005}
{Barklem}, P.~S., {Christlieb}, N., {Beers}, T.~C., {et~al.} 2005, \aap, 439,
  129

\bibitem[{{Beers} \& {Christlieb}(2005)}]{beers2005}
{Beers}, T.~C., \& {Christlieb}, N. 2005, \araa, 43, 531

\bibitem[{{Beers} {et~al.}(1985){Beers}, {Preston}, \& {Shectman}}]{beers1985}
{Beers}, T.~C., {Preston}, G.~W., \& {Shectman}, S.~A. 1985, \aj, 90, 2089

\bibitem[{{Beers} {et~al.}(1992){Beers}, {Preston}, \& {Shectman}}]{beers1992}
---. 1992, \aj, 103, 1987

\bibitem[{{Bernstein} {et~al.}(2003){Bernstein}, {Shectman}, {Gunnels},
  {Mochnacki}, \& {Athey}}]{mike}
{Bernstein}, R., {Shectman}, S.~A., {Gunnels}, S.~M., {Mochnacki}, S., \&
  {Athey}, A.~E. 2003, in Society of Photo-Optical Instrumentation Engineers
  (SPIE) Conference Series, Vol. 4841, Society of Photo-Optical Instrumentation
  Engineers (SPIE) Conference Series, ed. {M.~Iye \& A.~F.~M.~Moorwood}, 1694

\bibitem[{{Bessell} {et~al.}(2015){Bessell}, {Collett}, {Keller}, {Frebel},
  {Heger}, {Casey}, {Masseron}, {Asplund}, {Jacobson}, {Lind}, {Marino},
  {Norris}, {Yong}, {Da Costa}, {Chan}, {Magic}, {Schmidt}, \&
  {Tisserand}}]{bessell2015}
{Bessell}, M., {Collett}, R., {Keller}, S., {et~al.} 2015, ArXiv e-prints,
  arXiv:1505.03756

\bibitem[{{Bonifacio} {et~al.}(2012){Bonifacio}, {Sbordone}, {Caffau},
  {Ludwig}, {Spite}, {Gonz{\'a}lez Hern{\'a}ndez}, \& {Behara}}]{bonifacio2012}
{Bonifacio}, P., {Sbordone}, L., {Caffau}, E., {et~al.} 2012, \aap, 542, A87

\bibitem[{{Bonifacio} {et~al.}(2015){Bonifacio}, {Caffau}, {Spite}, {Limongi},
  {Chieffi}, {Klessen}, {Fran{\c c}ois}, {Molaro}, {Ludwig}, {Zaggia}, {Spite},
  {Plez}, {Cayrel}, {Christlieb}, {Clark}, {Glover}, {Hammer}, {Koch},
  {Monaco}, {Sbordone}, \& {Steffen}}]{bonifacio2015}
{Bonifacio}, P., {Caffau}, E., {Spite}, M., {et~al.} 2015, ArXiv e-prints,
  arXiv:1504.05963

\bibitem[{{Caffau} {et~al.}(2011{\natexlab{a}}){Caffau}, {Bonifacio}, {Fran{\c
  c}ois}, {Sbordone}, {Monaco}, {Spite}, {Spite}, {Ludwig}, {Cayrel}, {Zaggia},
  {Hammer}, {Randich}, {Molaro}, \& {Hill}}]{caffau2011b}
{Caffau}, E., {Bonifacio}, P., {Fran{\c c}ois}, P., {et~al.}
  2011{\natexlab{a}}, \nat, 477, 67

\bibitem[{{Caffau} {et~al.}(2011{\natexlab{b}}){Caffau}, {Bonifacio}, {Fran{\c
  c}ois}, {Spite}, {Spite}, {Zaggia}, {Ludwig}, {Monaco}, {Sbordone}, {Cayrel},
  {Hammer}, {Randich}, {Hill}, \& {Molaro}}]{caffau2011a}
---. 2011{\natexlab{b}}, \aap, 534, A4

\bibitem[{{Castelli} \& {Kurucz}(2004)}]{castelli2004}
{Castelli}, F., \& {Kurucz}, R.~L. 2004, ArXiv Astrophysics e-prints,
  arXiv:astro-ph/0405087

\bibitem[{{Christlieb} {et~al.}(2008){Christlieb}, {Sch{\"o}rck}, {Frebel},
  {Beers}, {Wisotzki}, \& {Reimers}}]{christlieb2008}
{Christlieb}, N., {Sch{\"o}rck}, T., {Frebel}, A., {et~al.} 2008, \aap, 484,
  721

\bibitem[{{Christlieb} {et~al.}(2002){Christlieb}, {Bessell}, {Beers},
  {Gustafsson}, {Korn}, {Barklem}, {Karlsson}, {Mizuno-Wiedner}, \&
  {Rossi}}]{christlieb2002}
{Christlieb}, N., {Bessell}, M.~S., {Beers}, T.~C., {et~al.} 2002, \nat, 419,
  904

\bibitem[{{Demarque} {et~al.}(2004){Demarque}, {Woo}, {Kim}, \&
  {Yi}}]{demarque2004}
{Demarque}, P., {Woo}, J.-H., {Kim}, Y.-C., \& {Yi}, S.~K. 2004, \apjs, 155,
  667

\bibitem[Frebel et al.(2015)]{frebel2015} Frebel, A., Chiti, A., 
Ji, A.~P., Jacobson, H.~R., \& Placco, V.~M.\ 2015, ArXiv e-prints,
arXiv:1507.01973 

\bibitem[{{Frebel} {et~al.}(2013){Frebel}, {Casey}, {Jacobson}, \&
  {Yu}}]{frebel2013}
{Frebel}, A., {Casey}, A.~R., {Jacobson}, H.~R., \& {Yu}, Q. 2013, \apj, 769,
  57

\bibitem[{{Frebel} \& {Norris}(2013)}]{frebel2011}
{Frebel}, A., \& {Norris}, J.~E. 2013, {Metal-Poor Stars and the Chemical
  Enrichment of the Universe} (Published), 55

\bibitem[{{Frebel} {et~al.}(2007){Frebel}, {Norris}, {Aoki}, {Honda},
  {Bessell}, {Takada-Hidai}, {Beers}, \& {Christlieb}}]{frebel2007}
{Frebel}, A., {Norris}, J.~E., {Aoki}, W., {et~al.} 2007, \apj, 658, 534

\bibitem[{{Frebel} {et~al.}(2014){Frebel}, {Simon}, \& {Kirby}}]{frebel2014}
{Frebel}, A., {Simon}, J.~D., \& {Kirby}, E.~N. 2014, \apj, 786, 74

\bibitem[{{Frebel} {et~al.}(2005){Frebel}, {Aoki}, {Christlieb}, {Ando},
  {Asplund}, {Barklem}, {Beers}, {Eriksson}, {Fechner}, {Fujimoto}, {Honda},
  {Kajino}, {Minezaki}, {Nomoto}, {Norris}, \& {Ryan}}]{frebel2005}
{Frebel}, A., {Aoki}, W., {Christlieb}, N., {et~al.} 2005, \nat, 434, 871

\bibitem[{{Frebel} {et~al.}(2006){Frebel}, {Christlieb}, {Norris}, {Beers},
  {Bessell}, {Rhee}, {Fechner}, {Marsteller}, {Rossi}, {Thom}, {Wisotzki}, \&
  {Reimers}}]{frebel2006}
{Frebel}, A., {Christlieb}, N., {Norris}, J.~E., {et~al.} 2006, \apj, 652, 1585

\bibitem[{{Hansen} {et~al.}(2014){Hansen}, {Hansen}, {Christlieb}, {Yong},
  {Bessell}, {Garc{\'{\i}}a P{\'e}rez}, {Beers}, {Placco}, {Frebel}, {Norris},
  \& {Asplund}}]{hansen2014}
{Hansen}, T., {Hansen}, C.~J., {Christlieb}, N., {et~al.} 2014, \apj, 787, 162

\bibitem[{{Hansen} {et~al.}(2015){Hansen}, {Hansen}, {Christlieb}, {Beers},
  {Yong}, {Bessell}, {Frebel}, {Garcia Perez}, {Placco}, {Norris}, \&
  {Asplund}}]{hansen2015}
---. 2015, ArXiv e-prints, arXiv:1506.00579

\bibitem[{{Heger} \& {Woosley}(2002)}]{heger2002}
{Heger}, A., \& {Woosley}, S.~E. 2002, \apj, 567, 532

\bibitem[{{Heger} \& {Woosley}(2010)}]{heger2010}
---. 2010, \apj, 724, 341

\bibitem[{{Ishigaki} {et~al.}(2014){Ishigaki}, {Tominaga}, {Kobayashi}, \&
  {Nomoto}}]{ishigaki2014}
{Ishigaki}, M.~N., {Tominaga}, N., {Kobayashi}, C., \& {Nomoto}, K. 2014,
  \apjl, 792, L32

\bibitem[{{Iwamoto} {et~al.}(2005){Iwamoto}, {Umeda}, {Tominaga}, {Nomoto}, \&
  {Maeda}}]{iwamoto2005}
{Iwamoto}, N., {Umeda}, H., {Tominaga}, N., {Nomoto}, K., \& {Maeda}, K. 2005,
  Science, 309, 451

\bibitem[{{Joggerst} {et~al.}(2009){Joggerst}, {Woosley}, \&
  {Heger}}]{joggerst2009}
{Joggerst}, C.~C., {Woosley}, S.~E., \& {Heger}, A. 2009, \apj, 693, 1780

\bibitem[{{Keller} {et~al.}(2014){Keller}, {Bessell}, {Frebel}, {Casey},
  {Asplund}, {Jacobson}, {Lind}, {Norris}, {Yong}, {Heger}, {Magic}, {da
  Costa}, {Schmidt}, \& {Tisserand}}]{keller2014}
{Keller}, S.~C., {Bessell}, M.~S., {Frebel}, A., {et~al.} 2014, \nat, 506, 463

\bibitem[{{Kelson}(2003)}]{kelson2003}
{Kelson}, D.~D. 2003, \pasp, 115, 688

\bibitem[{{Kupka} {et~al.}(1999){Kupka}, {Piskunov}, {Ryabchikova}, {Stempels},
  \& {Weiss}}]{vald}
{Kupka}, F., {Piskunov}, N., {Ryabchikova}, T.~A., {Stempels}, H.~C., \&
  {Weiss}, W.~W. 1999, \aaps, 138, 119

\bibitem[{{Lee} {et~al.}(2008{\natexlab{a}}){Lee}, {Beers}, {Sivarani},
  {Allende Prieto}, {Koesterke}, {Wilhelm}, {Re Fiorentin}, {Bailer-Jones},
  {Norris}, {Rockosi}, {Yanny}, {Newberg}, {Covey}, {Zhang}, \&
  {Luo}}]{lee2008a}
{Lee}, Y.~S., {Beers}, T.~C., {Sivarani}, T., {et~al.} 2008{\natexlab{a}}, \aj,
  136, 2022

\bibitem[{{Lee} {et~al.}(2008{\natexlab{b}}){Lee}, {Beers}, {Sivarani},
  {Johnson}, {An}, {Wilhelm}, {Allende Prieto}, {Koesterke}, {Re Fiorentin},
  {Bailer-Jones}, {Norris}, {Yanny}, {Rockosi}, {Newberg}, {Cudworth}, \&
  {Pan}}]{lee2008b}
---. 2008{\natexlab{b}}, \aj, 136, 2050

\bibitem[{{Lee} {et~al.}(2011){Lee}, {Beers}, {Allende Prieto}, {Lai},
  {Rockosi}, {Morrison}, {Johnson}, {An}, {Sivarani}, \& {Yanny}}]{lee2011}
{Lee}, Y.~S., {Beers}, T.~C., {Allende Prieto}, C., {et~al.} 2011, \aj, 141, 90

\bibitem[{{Lee} {et~al.}(2013){Lee}, {Beers}, {Masseron}, {Plez}, {Rockosi},
  {Sobeck}, {Yanny}, {Lucatello}, {Sivarani}, {Placco}, \& {Carollo}}]{lee2013}
{Lee}, Y.~S., {Beers}, T.~C., {Masseron}, T., {et~al.} 2013, \aj, 146, 132

\bibitem[{{Li} {et~al.}(2010){Li}, {Christlieb}, {Sch{\"o}rck}, {Norris},
  {Bessell}, {Yong}, {Beers}, {Lee}, \& {Frebel}}]{li2010}
{Li}, H.~N., {Christlieb}, N., {Sch{\"o}rck}, T., {et~al.} 2010, \aap, 521, 10

\bibitem[Lind et al.(2009)]{lind2009} Lind, K., Asplund, M., \& Barklem, P.~S.\ 2009,
\aap, 503, 541

\bibitem[{{Masseron} {et~al.}(2012){Masseron}, {Johnson}, {Lucatello},
  {Karakas}, {Plez}, {Beers}, \& {Christlieb}}]{masseron2012}
{Masseron}, T., {Johnson}, J.~A., {Lucatello}, S., {et~al.} 2012, \apj, 751, 14

\bibitem[{{Mel{\'e}ndez} {et~al.}(2010){Mel{\'e}ndez}, {Casagrande},
  {Ram{\'{\i}}rez}, {Asplund}, \& {Schuster}}]{melendez2010}
{Mel{\'e}ndez}, J., {Casagrande}, L., {Ram{\'{\i}}rez}, I., {Asplund}, M., \&
  {Schuster}, W.~J. 2010, \aap, 515, L3

\bibitem[{{Meynet} {et~al.}(2006){Meynet}, {Ekstr{\"o}m}, \&
  {Maeder}}]{meynet2006}
{Meynet}, G., {Ekstr{\"o}m}, S., \& {Maeder}, A. 2006, \aap, 447, 623

\bibitem[{{Meynet} {et~al.}(2010){Meynet}, {Hirschi}, {Ekstrom}, {Maeder},
  {Georgy}, {Eggenberger}, \& {Chiappini}}]{meynet2010}
{Meynet}, G., {Hirschi}, R., {Ekstrom}, S., {et~al.} 2010, \aap, 521, A30

\bibitem[{{Nomoto} {et~al.}(2013){Nomoto}, {Kobayashi}, \&
  {Tominaga}}]{nomoto2013}
{Nomoto}, K., {Kobayashi}, C., \& {Tominaga}, N. 2013, \araa, 51, 457

\bibitem[{{Nomoto} {et~al.}(2006){Nomoto}, {Tominaga}, {Umeda}, {Kobayashi}, \&
  {Maeda}}]{nomoto2006}
{Nomoto}, K., {Tominaga}, N., {Umeda}, H., {Kobayashi}, C., \& {Maeda}, K.
  2006, Nuclear Physics A, 777, 424

\bibitem[{{Placco} {et~al.}(2014{\natexlab{a}}){Placco}, {Frebel}, {Beers},
  {Christlieb}, {Lee}, {Kennedy}, {Rossi}, \& {Santucci}}]{placco2014}
{Placco}, V.~M., {Frebel}, A., {Beers}, T.~C., {et~al.} 2014{\natexlab{a}},
  \apj, 781, 40

\bibitem[{{Placco} {et~al.}(2013){Placco}, {Frebel}, {Beers}, {Karakas},
  {Kennedy}, {Rossi}, {Christlieb}, \& {Stancliffe}}]{placco2013}
---. 2013, \apj, 770, 104

\bibitem[{{Placco} {et~al.}(2014{\natexlab{b}}){Placco}, {Frebel}, {Beers}, \&
  {Stancliffe}}]{placco2014c}
{Placco}, V.~M., {Frebel}, A., {Beers}, T.~C., \& {Stancliffe}, R.~J.
  2014{\natexlab{b}}, \apj, 797, 21

\bibitem[{{Rockosi} {et~al.}(2015){Rockosi}, others, others,
  {et~al.}}]{rokosi2015}
{Rockosi}, C., others, others, {et~al.} 2015, \apj, in preparation

\bibitem[{{Roederer}(2013)}]{roederer2013}
{Roederer}, I.~U. 2013, \aj, 145, 26

\bibitem[Schneider et al.(2012)]{schneider2012} Schneider, R., 
Omukai, K., Limongi, M., et al.\ 2012, \mnras, 423, L60

\bibitem[{{Sch{\"o}rck} {et~al.}(2009){Sch{\"o}rck}, {Christlieb}, {Cohen},
  {Beers}, {Shectman}, {McWilliam}, {Bessell}, {Norris}, \&
  {Mel{\'e}ndez}}]{schorck2009}
{Sch{\"o}rck}, T., {Christlieb}, N., {Cohen}, J.~G., {et~al.} 2009, \aap, 507,
  817

\bibitem[{{Smolinski} {et~al.}(2011){Smolinski}, {Lee}, {Beers}, {An},
  {Bickerton}, {Johnson}, {Loomis}, {Rockosi}, {Sivarani}, \&
  {Yanny}}]{smolin2011}
{Smolinski}, J.~P., {Lee}, Y.~S., {Beers}, T.~C., {et~al.} 2011, \aj, 141, 89

\bibitem[{{Sneden}(1973)}]{sneden1973}
{Sneden}, C.~A. 1973, PhD thesis, The University of Texas at Austin.

\bibitem[{{Sobeck} {et~al.}(2011){Sobeck}, {Kraft}, {Sneden}, {Preston},
  {Cowan}, {Smith}, {Thompson}, {Shectman}, \& {Burley}}]{sobeck2011}
{Sobeck}, J.~S., {Kraft}, R.~P., {Sneden}, C., {et~al.} 2011, \aj, 141, 175

\bibitem[{{Spite} \& {Spite}(1982)}]{spite1982}
{Spite}, F., \& {Spite}, M. 1982, \aap, 115, 357

\bibitem[{{Spite} {et~al.}(2013){Spite}, {Caffau}, {Bonifacio}, {Spite},
  {Ludwig}, {Plez}, \& {Christlieb}}]{spite2013}
{Spite}, M., {Caffau}, E., {Bonifacio}, P., {et~al.} 2013, \aap, 552, A107

\bibitem[{{Tominaga} {et~al.}(2014){Tominaga}, {Iwamoto}, \&
  {Nomoto}}]{tominaga2014}
{Tominaga}, N., {Iwamoto}, N., \& {Nomoto}, K. 2014, \apj, 785, 98

\bibitem[{{Yanny} {et~al.}(2009){Yanny}, {Rockosi}, {Newberg}, {Knapp},
  {Adelman-McCarthy}, {Alcorn}, {Allam}, {Allende Prieto}, {An}, {Anderson},
  {Anderson}, \& {Bailer-Jones}}]{yanny2009}
{Yanny}, B., {Rockosi}, C., {Newberg}, H.~J., {et~al.} 2009, \aj, 137, 4377

\bibitem[{{Yong} {et~al.}(2013{\natexlab{a}}){Yong}, {Norris}, {Bessell},
  {Christlieb}, {Asplund}, {Beers}, {Barklem}, {Frebel}, \& {Ryan}}]{yong2013}
{Yong}, D., {Norris}, J.~E., {Bessell}, M.~S., {et~al.} 2013{\natexlab{a}},
  \apj, 762, 26

\bibitem[{{Yong} {et~al.}(2013{\natexlab{b}}){Yong}, {Norris}, {Bessell},
  {Christlieb}, {Asplund}, {Beers}, {Barklem}, {Frebel}, \& {Ryan}}]{yong2013b}
---. 2013{\natexlab{b}}, \apj, 762, 27

\bibitem[{{York} {et~al.}(2000){York}, {Adelman}, {Anderson}, {Anderson},
  {Annis}, {Bahcall}, {Bakken}, {Barkhouser}, {Bastian}, {Berman}, {Boroski},
  {Bracker}, \& {Briegel}}]{york2000}
{York}, D.~G., {Adelman}, J., {Anderson}, Jr., J.~E., {et~al.} 2000, \aj, 120,
  1579

\end{thebibliography}

\clearpage

\ltab

\begin{deluxetable}{lrr}
\tablewidth{0pt}
\tabletypesize{\scriptsize}
\tablecaption{Observational Data \label{candlist}}
\tablehead{
\colhead{} &
\colhead{\emp\xx} &
\colhead{\ump\yy}}
\startdata
$\alpha$ (J2000)  &    13:22:50.6 &    12:04:41.4 \\
$\delta$ (J2000)  & $+$01:23:43.0 & $+$12:01:11.5 \\
$g$ (mag)         &          16.3 &          16.4 \\
$g-r$             &          0.50 &          0.34 \\
\hline
\multicolumn{3}{c}{High Resolution -- Magellan/MIKE} \\
\hline
Date              &    2013 05 30 &    2013 05 31 \\
UT                &      02:35:55 &      00:09:39 \\
Exptime (s)       &          3600 &          5400 \\
v$_{r}$(km/s)     &         87.84 &         83.00
\enddata
\tablenotetext{a}{SDSS ID: 3307-54970-529}
\tablenotetext{b}{SDSS ID: 3214-54866-429}
\end{deluxetable}

\begin{deluxetable}{ccccccccc}
\tablewidth{0pc}
\tabletypesize{\scriptsize}
\tablecaption{Derived Stellar Parameters \label{obstable}}
\tablehead{&
\multicolumn{3}{c}{Medium Resolution}&&
\multicolumn{4}{c}{High Resolution}\\
\cline{2-4} \cline{6-9}\\
\colhead{}&
\colhead{\teff{}(K)    }&
\colhead{\logg{}(cgs)  }&
\colhead{\metal        }&
\colhead{              }&
\colhead{\teff{}(K)    }&
\colhead{\logg{}(cgs)  }&
\colhead{$\xi$(km/s)   }&
\colhead{\metal{}      }}
\startdata
\emp & 5466 (150) & 3.12 (0.35) &  $-$3.32 (0.20) && 5008 (100) & 1.95 (0.20) & 1.95 (0.20) & $-$3.64 (0.05) \\
\ump & 5894 (150) & 2.66 (0.35) &  $-$3.41 (0.20) && 5467 (100) & 3.20 (0.20) & 1.50 (0.20) & $-$4.34 (0.05) \\
\enddata
\end{deluxetable}

\begin{deluxetable}{lrrrrrrrr}
\tabletypesize{\scriptsize}
\tablewidth{0pc}
\tablecaption{\label{eqw} Equivalent-Width Measurements}
\tablehead{
\colhead{}&
\colhead{}&
\colhead{}&
\colhead{}&
\multicolumn{2}{c}{\emp}&
\colhead{}&
\multicolumn{2}{c}{\ump}\\
\cline{5-6} \cline{8-9} \\
\colhead{Ion}&
\colhead{$\lambda$}&
\colhead{$\chi$} &
\colhead{$\log\,gf$}&
\colhead{$W$}&
\colhead{$\log\epsilon$\,(X)}&
\colhead{}&
\colhead{$W$}&
\colhead{$\log\epsilon$\,(X)}\\
\colhead{}&
\colhead{({\AA})}&
\colhead{(eV)} &
\colhead{}&
\colhead{(m{\AA})}&
\colhead{}&
\colhead{}&
\colhead{(m{\AA})}&
\colhead{}}
\startdata
    Na I & 5889.950 & 0.00 &       0.108 &     62.8 &      2.30 &&     39.0 &      2.28 \\
    Na I & 5895.924 & 0.00 &    $-$0.194 &     61.9 &      2.58 &&     30.3 &      2.41 \\
    Mg I & 3829.355 & 2.71 &    $-$0.208 &     98.5 &      4.01 &&  \nodata &   \nodata \\
    Mg I & 3832.304 & 2.71 &       0.270 &    132.3 &      4.22 &&     89.6 &      3.62 \\
    Mg I & 4167.271 & 4.35 &    $-$0.710 &      9.1 &      4.35 &&  \nodata &   \nodata \\
    Mg I & 4702.990 & 4.33 &    $-$0.380 &     13.6 &      4.18 &&  \nodata &   \nodata \\
    Mg I & 5172.684 & 2.71 &    $-$0.450 &    105.0 &      4.19 &&     63.5 &      3.78 \\
    Mg I & 5183.604 & 2.72 &    $-$0.239 &    117.2 &      4.23 &&     64.1 &      3.59 \\
    Mg I & 5528.405 & 4.34 &    $-$0.498 &     12.9 &      4.27 &&  \nodata &   \nodata \\
    Al I & 3961.520 & 0.01 &    $-$0.340 &     37.6 &      1.49 &&  \nodata &   \nodata \\
    Si I & 3905.523 & 1.91 &    $-$1.092 &    106.5 &      4.14 &&     71.8 &      3.88 \\
    Ca I & 4226.730 & 0.00 &       0.244 &    103.8 &      2.52 &&     77.1 &      2.42 \\
    Ca I & 4283.010 & 1.89 &    $-$0.224 &     16.2 &      3.17 &&  \nodata &   \nodata \\
    Ca I & 4434.960 & 1.89 &    $-$0.010 &     16.8 &      2.97 &&  \nodata &   \nodata \\
    Ca I & 4454.780 & 1.90 &       0.260 &     24.6 &      2.92 &&  \nodata &   \nodata \\
    Ca I & 6162.170 & 1.90 &    $-$0.089 &     18.8 &      3.06 &&  \nodata &   \nodata \\
    Ca I & 6439.070 & 2.52 &       0.470 &     12.0 &      2.96 &&  \nodata &   \nodata \\
   Ca II & 3933.663 & 0.00 &       0.105 &  \nodata &   \nodata &&    728.3 &      2.35 \\
   Sc II & 4246.820 & 0.32 &       0.240 &     47.7 &   $-$0.86 &&     13.6 &   $-$0.78 \\
   Sc II & 4314.083 & 0.62 &    $-$0.100 &     20.4 &   $-$0.73 &&  \nodata &   \nodata \\
   Ti II & 3759.291 & 0.61 &       0.280 &    124.6 &      1.80 &&  \nodata &   \nodata \\
   Ti II & 3761.320 & 0.57 &       0.180 &    116.8 &      1.67 &&  \nodata &   \nodata \\
   Ti II & 3813.394 & 0.61 &    $-$2.020 &     20.9 &      1.62 &&  \nodata &   \nodata \\
   Ti II & 3913.461 & 1.12 &    $-$0.420 &     48.6 &      1.17 &&  \nodata &   \nodata \\
   Ti II & 4012.396 & 0.57 &    $-$1.750 &     44.6 &      1.78 &&  \nodata &   \nodata \\
   Ti II & 4025.120 & 0.61 &    $-$1.980 &     21.8 &      1.58 &&  \nodata &   \nodata \\
   Ti II & 4290.219 & 1.16 &    $-$0.930 &     31.4 &      1.36 &&  \nodata &   \nodata \\
   Ti II & 4337.914 & 1.08 &    $-$0.960 &     37.0 &      1.41 &&  \nodata &   \nodata \\
   Ti II & 4395.031 & 1.08 &    $-$0.540 &     52.4 &      1.26 &&  \nodata &   \nodata \\
   Ti II & 4399.765 & 1.24 &    $-$1.190 &     22.9 &      1.51 &&  \nodata &   \nodata \\
   Ti II & 4443.801 & 1.08 &    $-$0.720 &     45.0 &      1.30 &&  \nodata &   \nodata \\
   Ti II & 4450.482 & 1.08 &    $-$1.520 &     14.8 &      1.42 &&  \nodata &   \nodata \\
   Ti II & 4468.517 & 1.13 &    $-$0.600 &     41.0 &      1.17 &&     19.0 &      1.48 \\
   Ti II & 4501.270 & 1.12 &    $-$0.770 &     37.7 &      1.26 &&  \nodata &   \nodata \\
   Ti II & 4533.960 & 1.24 &    $-$0.530 &     38.3 &      1.17 &&  \nodata &   \nodata \\
   Ti II & 4563.770 & 1.22 &    $-$0.960 &     30.1 &      1.41 &&  \nodata &   \nodata \\
   Ti II & 4571.971 & 1.57 &    $-$0.320 &     32.8 &      1.23 &&  \nodata &   \nodata \\
    Cr I & 4254.332 & 0.00 &    $-$0.114 &     45.4 &      1.33 &&  \nodata &   \nodata \\
    Cr I & 4274.800 & 0.00 &    $-$0.220 &     49.8 &      1.52 &&  \nodata &   \nodata \\
    Cr I & 4289.720 & 0.00 &    $-$0.370 &     26.5 &      1.21 &&  \nodata &   \nodata \\
    Cr I & 5206.040 & 0.94 &       0.020 &     23.3 &      1.76 &&  \nodata &   \nodata \\
    Cr I & 5208.419 & 0.94 &       0.160 &     15.0 &      1.39 &&     16.8 &      1.89 \\
    Mn I & 4030.753 & 0.00 &    $-$0.480 &     58.8 &      1.28 &&  \nodata &   \nodata \\
    Mn I & 4033.062 & 0.00 &    $-$0.618 &     40.7 &      1.07 &&  \nodata &   \nodata \\
    Mn I & 4034.483 & 0.00 &    $-$0.811 &     45.2 &      1.34 &&  \nodata &   \nodata \\
    Fe I & 3727.619 & 0.96 &    $-$0.609 &     97.6 &      3.81 &&  \nodata &   \nodata \\
    Fe I & 3743.362 & 0.99 &    $-$0.790 &     96.8 &      4.00 &&  \nodata &   \nodata \\
    Fe I & 3753.611 & 2.18 &    $-$0.890 &     23.6 &      3.70 &&  \nodata &   \nodata \\
    Fe I & 3763.789 & 0.99 &    $-$0.221 &    109.3 &      3.76 &&  \nodata &   \nodata \\
    Fe I & 3765.539 & 3.24 &       0.482 &     41.0 &      3.89 &&  \nodata &   \nodata \\
    Fe I & 3767.192 & 1.01 &    $-$0.390 &    109.9 &      3.96 &&     58.4 &      3.23 \\
    Fe I & 3786.677 & 1.01 &    $-$2.185 &     41.0 &      4.04 &&  \nodata &   \nodata \\
    Fe I & 3787.880 & 1.01 &    $-$0.838 &     87.0 &      3.78 &&  \nodata &   \nodata \\
    Fe I & 3815.840 & 1.48 &       0.237 &    107.1 &      3.76 &&     61.6 &      3.17 \\
    Fe I & 3820.425 & 0.86 &       0.157 &  \nodata &   \nodata &&     77.1 &      3.05 \\
    Fe I & 3824.444 & 0.00 &    $-$1.360 &    118.1 &      3.97 &&  \nodata &   \nodata \\
    Fe I & 3825.881 & 0.91 &    $-$0.024 &  \nodata &   \nodata &&     76.7 &      3.27 \\
    Fe I & 3827.823 & 1.56 &       0.094 &     98.2 &      3.77 &&  \nodata &   \nodata \\
    Fe I & 3840.438 & 0.99 &    $-$0.497 &    103.1 &      3.85 &&  \nodata &   \nodata \\
    Fe I & 3841.048 & 1.61 &    $-$0.044 &     88.9 &      3.69 &&  \nodata &   \nodata \\
    Fe I & 3846.800 & 3.25 &    $-$0.020 &     19.4 &      3.90 &&  \nodata &   \nodata \\
    Fe I & 3849.967 & 1.01 &    $-$0.863 &     96.4 &      4.04 &&  \nodata &   \nodata \\
    Fe I & 3850.818 & 0.99 &    $-$1.745 &     59.4 &      3.94 &&  \nodata &   \nodata \\
    Fe I & 3856.372 & 0.05 &    $-$1.280 &    120.0 &      3.98 &&  \nodata &   \nodata \\
    Fe I & 3859.911 & 0.00 &    $-$0.710 &  \nodata &   \nodata &&     85.8 &      3.27 \\
    Fe I & 3865.523 & 1.01 &    $-$0.950 &     80.2 &      3.67 &&  \nodata &   \nodata \\
    Fe I & 3878.018 & 0.96 &    $-$0.896 &     85.8 &      3.72 &&  \nodata &   \nodata \\
    Fe I & 3878.573 & 0.09 &    $-$1.380 &    110.8 &      3.90 &&     57.5 &      3.21 \\
    Fe I & 3886.282 & 0.05 &    $-$1.080 &  \nodata &   \nodata &&     65.3 &      3.08 \\
    Fe I & 3887.048 & 0.91 &    $-$1.140 &  \nodata &   \nodata &&     32.5 &      3.24 \\
    Fe I & 3895.656 & 0.11 &    $-$1.668 &     95.2 &      3.78 &&     42.8 &      3.15 \\
    Fe I & 3899.707 & 0.09 &    $-$1.515 &    106.2 &      3.90 &&  \nodata &   \nodata \\
    Fe I & 3902.946 & 1.56 &    $-$0.442 &     86.8 &      3.96 &&  \nodata &   \nodata \\
    Fe I & 3917.181 & 0.99 &    $-$2.155 &     35.3 &      3.85 &&  \nodata &   \nodata \\
    Fe I & 3920.258 & 0.12 &    $-$1.734 &     98.1 &      3.92 &&     41.4 &      3.19 \\
    Fe I & 3922.912 & 0.05 &    $-$1.626 &    107.8 &      4.01 &&     45.4 &      3.10 \\
    Fe I & 3940.878 & 0.96 &    $-$2.600 &     26.4 &      4.07 &&  \nodata &   \nodata \\
    Fe I & 3949.953 & 2.18 &    $-$1.251 &     13.8 &      3.75 &&  \nodata &   \nodata \\
    Fe I & 3977.741 & 2.20 &    $-$1.120 &     22.3 &      3.90 &&  \nodata &   \nodata \\
    Fe I & 4005.242 & 1.56 &    $-$0.583 &     79.1 &      3.87 &&  \nodata &   \nodata \\
    Fe I & 4045.812 & 1.49 &       0.284 &    106.8 &      3.66 &&     59.9 &      3.07 \\
    Fe I & 4063.594 & 1.56 &       0.062 &     99.0 &      3.75 &&     46.1 &      3.02 \\
    Fe I & 4067.978 & 3.21 &    $-$0.470 &     11.8 &      4.04 &&  \nodata &   \nodata \\
    Fe I & 4071.738 & 1.61 &    $-$0.008 &    101.7 &      3.94 &&    423.0 &      3.07 \\
    Fe I & 4132.058 & 1.61 &    $-$0.675 &     74.5 &      3.88 &&  \nodata &   \nodata \\
    Fe I & 4134.678 & 2.83 &    $-$0.649 &     11.3 &      3.77 &&  \nodata &   \nodata \\
    Fe I & 4143.868 & 1.56 &    $-$0.511 &     75.7 &      3.68 &&     28.3 &      3.18 \\
    Fe I & 4147.669 & 1.48 &    $-$2.071 &     15.4 &      3.82 &&  \nodata &   \nodata \\
    Fe I & 4181.755 & 2.83 &    $-$0.371 &     23.6 &      3.88 &&  \nodata &   \nodata \\
    Fe I & 4187.039 & 2.45 &    $-$0.514 &     31.9 &      3.77 &&  \nodata &   \nodata \\
    Fe I & 4187.795 & 2.42 &    $-$0.510 &     36.2 &      3.83 &&  \nodata &   \nodata \\
    Fe I & 4191.430 & 2.47 &    $-$0.666 &     26.7 &      3.84 &&  \nodata &   \nodata \\
    Fe I & 4199.095 & 3.05 &       0.156 &     26.9 &      3.67 &&  \nodata &   \nodata \\
    Fe I & 4202.029 & 1.49 &    $-$0.689 &     71.1 &      3.66 &&  \nodata &   \nodata \\
    Fe I & 4216.184 & 0.00 &    $-$3.357 &     36.0 &      3.90 &&  \nodata &   \nodata \\
    Fe I & 4222.213 & 2.45 &    $-$0.914 &     24.2 &      4.00 &&  \nodata &   \nodata \\
    Fe I & 4233.603 & 2.48 &    $-$0.579 &     29.0 &      3.81 &&  \nodata &   \nodata \\
    Fe I & 4250.119 & 2.47 &    $-$0.380 &     32.3 &      3.67 &&  \nodata &   \nodata \\
    Fe I & 4250.787 & 1.56 &    $-$0.713 &     79.6 &      3.97 &&     17.3 &      3.07 \\
    Fe I & 4260.474 & 2.40 &       0.077 &     69.0 &      3.86 &&  \nodata &   \nodata \\
    Fe I & 4271.154 & 2.45 &    $-$0.337 &     40.5 &      3.77 &&  \nodata &   \nodata \\
    Fe I & 4271.760 & 1.49 &    $-$0.173 &     91.8 &      3.66 &&     50.9 &      3.27 \\
    Fe I & 4282.403 & 2.18 &    $-$0.779 &     29.5 &      3.68 &&  \nodata &   \nodata \\
    Fe I & 4325.762 & 1.61 &       0.006 &     94.2 &      3.67 &&     52.7 &      3.26 \\
    Fe I & 4337.046 & 1.56 &    $-$1.695 &     34.6 &      4.00 &&  \nodata &   \nodata \\
    Fe I & 4375.930 & 0.00 &    $-$3.005 &     46.8 &      3.73 &&  \nodata &   \nodata \\
    Fe I & 4383.545 & 1.48 &       0.200 &    115.6 &      3.86 &&     58.4 &      3.07 \\
    Fe I & 4404.750 & 1.56 &    $-$0.147 &     92.6 &      3.71 &&     45.1 &      3.18 \\
    Fe I & 4415.122 & 1.61 &    $-$0.621 &     74.6 &      3.78 &&  \nodata &   \nodata \\
    Fe I & 4427.310 & 0.05 &    $-$2.924 &     49.3 &      3.75 &&  \nodata &   \nodata \\
    Fe I & 4447.717 & 2.22 &    $-$1.339 &     15.2 &      3.90 &&  \nodata &   \nodata \\
    Fe I & 4459.118 & 2.18 &    $-$1.279 &     22.4 &      4.00 &&  \nodata &   \nodata \\
    Fe I & 4461.653 & 0.09 &    $-$3.194 &     38.1 &      3.85 &&  \nodata &   \nodata \\
    Fe I & 4476.019 & 2.85 &    $-$0.820 &     11.3 &      3.94 &&  \nodata &   \nodata \\
    Fe I & 4489.739 & 0.12 &    $-$3.899 &     14.4 &      4.02 &&  \nodata &   \nodata \\
    Fe I & 4494.563 & 2.20 &    $-$1.143 &     23.7 &      3.91 &&  \nodata &   \nodata \\
    Fe I & 4528.614 & 2.18 &    $-$0.822 &     41.5 &      3.94 &&  \nodata &   \nodata \\
    Fe I & 4531.148 & 1.48 &    $-$2.101 &     18.3 &      3.91 &&  \nodata &   \nodata \\
    Fe I & 4871.318 & 2.87 &    $-$0.362 &     20.5 &      3.79 &&  \nodata &   \nodata \\
    Fe I & 4872.137 & 2.88 &    $-$0.567 &     12.0 &      3.73 &&  \nodata &   \nodata \\
    Fe I & 4890.755 & 2.88 &    $-$0.394 &     21.6 &      3.86 &&  \nodata &   \nodata \\
    Fe I & 4891.492 & 2.85 &    $-$0.111 &     29.3 &      3.73 &&  \nodata &   \nodata \\
    Fe I & 4918.994 & 2.85 &    $-$0.342 &     19.3 &      3.71 &&  \nodata &   \nodata \\
    Fe I & 4920.503 & 2.83 &       0.068 &     36.5 &      3.67 &&  \nodata &   \nodata \\
    Fe I & 5006.119 & 2.83 &    $-$0.615 &     21.9 &      4.03 &&  \nodata &   \nodata \\
    Fe I & 5012.068 & 0.86 &    $-$2.642 &     32.6 &      4.04 &&  \nodata &   \nodata \\
    Fe I & 5051.634 & 0.92 &    $-$2.764 &     26.1 &      4.09 &&  \nodata &   \nodata \\
    Fe I & 5269.537 & 0.86 &    $-$1.333 &     98.2 &      4.11 &&     32.6 &      3.29 \\
    Fe I & 5328.039 & 0.92 &    $-$1.466 &     90.3 &      4.12 &&  \nodata &   \nodata \\
    Fe I & 5328.531 & 1.56 &    $-$1.850 &     18.0 &      3.70 &&  \nodata &   \nodata \\
    Fe I & 5371.489 & 0.96 &    $-$1.644 &     80.7 &      4.10 &&  \nodata &   \nodata \\
    Fe I & 5397.128 & 0.92 &    $-$1.982 &     65.6 &      4.06 &&  \nodata &   \nodata \\
    Fe I & 5405.775 & 0.99 &    $-$1.852 &     58.9 &      3.88 &&  \nodata &   \nodata \\
    Fe I & 5429.696 & 0.96 &    $-$1.881 &     62.0 &      3.94 &&  \nodata &   \nodata \\
    Fe I & 5434.524 & 1.01 &    $-$2.126 &     36.6 &      3.76 &&  \nodata &   \nodata \\
    Fe I & 5446.917 & 0.99 &    $-$1.910 &     54.9 &      3.86 &&  \nodata &   \nodata \\
    Fe I & 5455.609 & 1.01 &    $-$2.090 &     45.2 &      3.88 &&  \nodata &   \nodata \\
    Fe I & 5506.779 & 0.99 &    $-$2.789 &     16.0 &      3.91 &&  \nodata &   \nodata \\
   Fe II & 4233.170 & 2.58 &    $-$1.970 &     27.2 &      3.85 &&  \nodata &   \nodata \\
   Fe II & 4522.630 & 2.84 &    $-$2.250 &      9.3 &      3.84 &&  \nodata &   \nodata \\
   Fe II & 4583.840 & 2.81 &    $-$1.930 &     22.0 &      3.93 &&  \nodata &   \nodata \\
   Fe II & 4923.930 & 2.89 &    $-$1.320 &     41.6 &      3.81 &&  \nodata &   \nodata \\
   Fe II & 5018.450 & 2.89 &    $-$1.220 &     49.2 &      3.85 &&  \nodata &   \nodata \\
    Co I & 3845.468 & 0.92 &       0.010 &     46.0 &      1.68 &&  \nodata &   \nodata \\
    Co I & 3873.120 & 0.43 &    $-$0.660 &     55.3 &      1.97 &&  \nodata &   \nodata \\
    Co I & 3881.869 & 0.58 &    $-$1.130 &     28.4 &      2.07 &&  \nodata &   \nodata \\
    Co I & 3995.306 & 0.92 &    $-$0.220 &     28.9 &      1.55 &&  \nodata &   \nodata \\
    Co I & 4121.318 & 0.92 &    $-$0.320 &     35.3 &      1.77 &&  \nodata &   \nodata \\
    Ni I & 3452.880 & 0.11 &    $-$0.900 &     76.8 &      2.45 &&  \nodata &   \nodata \\
    Ni I & 3483.770 & 0.28 &    $-$1.120 &     64.5 &      2.48 &&  \nodata &   \nodata \\
    Ni I & 3492.960 & 0.11 &    $-$0.265 &     92.8 &      2.30 &&  \nodata &   \nodata \\
    Ni I & 3519.770 & 0.28 &    $-$1.422 &     58.9 &      2.63 &&  \nodata &   \nodata \\
    Ni I & 3524.540 & 0.03 &       0.007 &    110.7 &      2.40 &&  \nodata &   \nodata \\
    Ni I & 3566.370 & 0.42 &    $-$0.251 &  \nodata &   \nodata &&     68.9 &      2.61 \\
    Ni I & 3597.710 & 0.21 &    $-$1.115 &     68.8 &      2.48 &&  \nodata &   \nodata \\
    Ni I & 3783.520 & 0.42 &    $-$1.420 &     54.2 &      2.56 &&  \nodata &   \nodata \\
    Ni I & 3807.140 & 0.42 &    $-$1.220 &     59.6 &      2.47 &&  \nodata &   \nodata \\
    Ni I & 3858.301 & 0.42 &    $-$0.951 &     71.5 &      2.47 &&  \nodata &   \nodata \\
    Ni I & 5476.900 & 1.83 &    $-$0.890 &     18.3 &      2.75 &&  \nodata &   \nodata \\
\enddata
\end{deluxetable}

\begin{deluxetable}{lrrrrrrrrrr}
\tablewidth{0pc}
\tabletypesize{\scriptsize}
\tablecaption{Abundances for Individual Species \label{abund}}
\tablehead{
&& \multicolumn{4}{c}{\emp} &  & \multicolumn{4}{c}{\ump} \\
 \cline{3-6} \cline{8-11}
Species & $\log\epsilon_{\odot}$\,(X) & $\log\epsilon$\,(X) & $\mbox{[X/Fe]}$ & $\sigma$ & $N$ &
                                      & $\log\epsilon$\,(X) & $\mbox{[X/Fe]}$ & $\sigma$ & $N$ }
\startdata
Li      & 1.05 &  \nodata &  \nodata    & \nodata & \nodata &&     1.70 &  \nodata    & 0.15    & 1       \\
C       & 8.43 &     5.28 &  $+$0.49\xx & 0.20    & 1       &&  $<$5.54 & $<+$1.45    & \nodata & 1       \\
Na I    & 6.24 &     2.44 &  $-$0.16    & 0.10    & 2       &&     2.34 &  $+$0.44    & 0.05    & 2       \\
Mg I    & 7.60 &     4.21 &  $+$0.25    & 0.05    & 7       &&     3.66 &  $+$0.40    & 0.05    & 3       \\
Al I    & 6.45 &     1.49 &  $-$1.32    & \nodata & 1       &&  \nodata &  \nodata    & \nodata & \nodata \\
Si I    & 7.51 &     4.14 &  $+$0.27    & \nodata & 1       &&     3.88 &  $+$0.71    & \nodata & 1       \\
Ca I    & 6.34 &     2.93 &  $+$0.23    & 0.08    & 6       &&     2.42 &  $+$0.42    & \nodata & 1       \\
Ca II   & 6.34 &  \nodata &  \nodata    & \nodata & \nodata &&     2.35 &  $+$0.35    & \nodata & 1       \\
Sc II   & 3.15 &  $-$0.79 &  $-$0.30    & 0.05    & 2       &&  $-$0.78 &  $+$0.41    & \nodata & 1       \\
Ti II   & 4.95 &     1.42 &  $+$0.11    & 0.05    & 17      &&     1.48 &  $+$0.87    & \nodata & 1       \\
Cr I    & 5.64 &     1.44 &  $-$0.56    & 0.08    & 5       &&     1.89 &  $+$0.59    & \nodata & 1       \\
Mn I    & 5.43 &     1.23 &  $-$0.56    & 0.07    & 3       &&  \nodata &  \nodata    & \nodata & \nodata \\
Fe I    & 7.50 &     3.86 &  $-$3.64\yy & 0.01    & 88      &&     3.16 &  $-$4.34\yy & 0.02    & 21      \\
Fe II   & 7.50 &     3.86 &  $-$3.64\yy & 0.02    & 5       &&  \nodata &  \nodata    & \nodata & \nodata \\
Co I    & 4.99 &     1.81 &  $+$0.46    & 0.08    & 6       &&  \nodata &  \nodata    & \nodata & \nodata \\
Ni I    & 6.22 &     2.50 &  $-$0.08    & 0.05    & 10      &&     2.61 &  $+$0.73    & \nodata & 1       \\
Sr II   & 2.87 &  $-$2.01 &  $-$1.24    & 0.20    & 2       && $<-$1.55 & $<-$0.08    & \nodata & 1       \\
Ba II   & 2.18 &  $-$2.76 &  $-$1.30    & 0.20    & 1       && $<-$1.54 & $<+$0.62    & \nodata & 1       \\
\enddata
\tablenotetext{a}{\cfe=$+$0.50 using corrections of \citet{placco2014c}.}
\tablenotetext{b}{$\mbox{[FeI/H]}$ and $\mbox{[FeII/H]}$ values}
\end{deluxetable}

\begin{deluxetable}{lrrrr}
\tabletypesize{\scriptsize}
\tablewidth{0pc}
\tablecaption{Example Systematic Abundance Uncertainties for \protect\emp \label{sys}}
\tablehead{
\colhead{Elem}&
\colhead{$\Delta$\teff}&
\colhead{$\Delta$\logg}&
\colhead{$\Delta\xi$}&
\colhead{$\sigma_{\rm tot}$}\\
\colhead{}&
\colhead{$+$150\,K}&
\colhead{$+$0.2 dex}&
\colhead{$+$0.2 km/s}&
\colhead{(dex)}}
\startdata
Na I  & 0.15 & $-$0.01 & $-$0.03 & 0.18 \\
Mg I  & 0.12 & $-$0.03 & $-$0.05 & 0.14 \\
Al I  & 0.15 & $-$0.02 & $-$0.02 & 0.18 \\
Si I  & 0.17 & $-$0.03 & $-$0.11 & 0.23 \\
Ca I  & 0.11 & $-$0.02 & $-$0.03 & 0.14 \\
Sc II & 0.10 &    0.06 & $-$0.02 & 0.14 \\
Ti II & 0.09 &    0.06 & $-$0.03 & 0.12 \\
Cr I  & 0.17 & $-$0.02 & $-$0.02 & 0.19 \\
Mn I  & 0.20 & $-$0.02 & $-$0.03 & 0.21 \\
Fe I  & 0.18 & $-$0.02 & $-$0.07 & 0.19 \\
Fe II & 0.03 &    0.07 & $-$0.02 & 0.09 \\
Co I  & 0.19 & $-$0.02 & $-$0.02 & 0.23 \\
Ni I  & 0.22 & $-$0.03 & $-$0.09 & 0.24 \\
Sr II & 0.11 &    0.07 & $-$0.03 & 0.16 \\
Ba II & 0.12 &    0.05 & $-$0.01 & 0.16
\enddata
\end{deluxetable}

\begin{deluxetable}{@{}rlrrrrrc@{}c@{}r@{}}
\tabletypesize{\scriptsize}
\tablewidth{0pc}
\tablecaption{Ultra Metal-Poor Stars from the Literature \label{umptab}}
\tablehead{
\colhead{ID}&
\colhead{Name}&
\colhead{\metal}&
\colhead{\cfe}&
\colhead{\eps{C}}&
\colhead{\eps{N}}&
\colhead{\eps{N}}&
\colhead{Progenitor}&
\colhead{Energy}&
\colhead{Reference}\\
\colhead{}&
\colhead{}&
\colhead{}&
\colhead{}&
\colhead{}&
\colhead{measured}&
\colhead{estimated\yy}&
\colhead{Mass (M$_{\odot}$)}&
\colhead{($\times 10^{51}$~erg)}&
\colhead{}}
\startdata
 1 &  SDSS~J2209$-$0028 &  $-$4.00 &  $+$2.56    &     6.99 &    \nodata  &     6.39 & 27.0 & ~0.3 & \citet{spite2013} \\
 2 &     HE~2139$-$5432 &  $-$4.02 &  $+$2.60\xx &     7.01 &     5.89    &  \nodata & 28.0 & ~0.6 & \citet{yong2013} \\
 3 &           G77$-$61 &  $-$4.03 &  $+$2.49    &     7.00 &     6.40    &  \nodata & 27.0 & ~0.3 & \citet{allen2012} \\
 4 &     CS~30336$-$049 &  $-$4.03 &  $+$0.09\xx &     4.85 &     4.70    &  \nodata & 21.5 & ~0.3 & \citet{yong2013} \\
 5 &     HE~1424$-$0241 &  $-$4.05 &  $+$0.63    &     5.01 &    \nodata  &     4.41 & 21.5 & ~0.3 & \citet{yong2013} \\
 6 &     HE~0057$-$5959 &  $-$4.08 &  $+$0.86    &     5.21 &     5.90    &  \nodata & 27.0 & ~0.3 & \citet{yong2013} \\
 7 &    SDSS~J0140$+$23 &  $-$4.09 &  $+$1.57    &     5.91 &    \nodata  &     5.31 & 27.0 & ~0.3 & \citet{yong2013} \\
 8 &     HE~2239$-$5019 &  $-$4.15 &  $+$1.80    & $<~$5.98 & $<~$6.38    &  \nodata & 15.0 & 10.0 & \citet{hansen2014} \\
 9 &     HE~1310$-$0536 &  $-$4.15 &  $+$2.53\xx &     6.72 & $<~$6.88    &  \nodata & 10.9 & ~0.3 & \citet{hansen2014} \\
10 &        CD$-$38~245 &  $-$4.15 &  $-$0.09\xx &     4.19 &     4.75    &  \nodata & 21.5 & ~0.3 & \citet{yong2013} \\
11 &  SDSS~J1204$+$1201 &  $-$4.34 & $<+$1.45    & $<~$5.54 &    \nodata  & $<~$4.94 & 10.6 & ~0.9 & This work \\
12 &     CS~22949$-$037 &  $-$4.38 &  $+$1.73\xx &     5.78 &     5.95    &  \nodata & 27.0 & ~0.3 & \citet{roederer2013} \\
13 &     HE~0233$-$0343 &  $-$4.68 &  $+$3.32    &     7.23 & $<~$5.95    &  \nodata & 11.9 & ~0.3 & \citet{hansen2014} \\
14 &     HE~0557$-$4840 &  $-$4.75 &  $+$1.66\xx &     5.30 & $<~$5.40    &  \nodata & 10.9 & ~0.6 & \citet{masseron2012} \\
15 &  SDSS~J1742$+$2531 &  $-$4.80 &  $+$3.63    &     7.26 &    \nodata  &     6.66 & 21.5 & ~0.3 & \citet{bonifacio2015} \\
16 &  SDSS~J1029$+$1729 &  $-$4.99 & $<+$0.70    & $<~$4.20 & $<~$3.10    &  \nodata & 10.6 & ~0.9 & \citet{caffau2011b} \\
17 &  SDSS~J1313$+$0019 &  $-$5.00 &  $+$2.96    &     6.39 &     6.29    &  \nodata & 27.0 & ~0.3 & \citet{frebel2015} \\
18 &  SDSS~J1035$+$0641 & $<-$5.07 & $>+$3.55    &     6.90 &    \nodata  &     6.30 & 23.0 & ~0.6 & \citet{bonifacio2015} \\
19 &     HE~0107$-$5240 &  $-$5.54 &  $+$2.69\xx &     5.58 &     3.80    &  \nodata & 20.5 & ~0.6 & \citet{christlieb2002} \\
20 &     HE~1327$-$2326 &  $-$5.65 &  $+$3.48    &     6.26 &     5.93    &  \nodata & 21.5 & ~0.3 & \citet{frebel2005} \\
21 &  SMSS~J0313$-$6708 & $<-$7.80 & $>+$5.39    &     6.02 & $<~$3.63    &  \nodata & 41.0 & ~1.2 & \citet{bessell2015} \\
\enddata
\tablenotetext{a}{Using corrections of \citet{placco2014c}.}
\tablenotetext{b}{Estimated from [C/N]=0. See text for details.}
\end{deluxetable}

\begin{deluxetable}{ll}
\tablewidth{0pc}
\tabletypesize{\scriptsize}
\tablecaption{Changes Applied to the Literature Data \label{casestab}}
\tablehead{
\colhead{Case \#}&
\colhead{Constraint}}
\startdata

(i)    & $\cfe-0.5$ \\
(ii)   & $\cfe-1.0$ \\
(iii)  & $\nfe-0.5$ \\
(iv)   & $\nfe-1.0$ \\
(v)    & $\nfe+0.5$ \\
(vi)   & $\nfe+1.0$ \\
(vii)  & $\nfe+2.0$ \\
(viii) & no carbon/nitrogen \\
(ix)   & no carbon \\
(x)    & no nitrogen \\

\enddata
\end{deluxetable}

\end{document}